\documentclass[a4paper,12pt]{article} 
\usepackage{graphicx}
\usepackage{amssymb}
\usepackage{amsmath}
\usepackage[colorlinks=true, pdfstartview=FitV, linkcolor=blue,  citecolor=blue, urlcolor=blue]{hyperref}  %liens hypertexte
\usepackage{float}
\usepackage{yfonts}
\usepackage[T1]{fontenc}

\graphicspath{ {images/} }

\newcommand{\dfr}{\widehat{d}} %fract deriv.: pas mal, op�rateur...
\newcommand{\Vsc}{{\widehat V}} % V rond
\newcommand{\Ds}{{\widetilde D}} % D rond
\def\l{\left}
\def\r{\right}
\def\beq{\begin{equation}}
\def\eeq{\end{equation}}
\def\d{\partial}

\def\beq{\begin{equation}}\def\eeq{\end{equation}}
\def\bea{\begin{eqnarray}}\def\eea{\end{eqnarray}}

\usepackage{parskip}

\begin{document}

\title{The physical principles underpinning self-organization in plants.}

\author{Philip Turner$^1$\footnote{ph.turner@napier.ac.uk}  and Laurent Nottale$^2$\footnote{laurent.nottale@obspm.fr}\footnote{\copyright 2016. This manuscript version is made available under the CC-BY-NC-ND 4.0 license http://creativecommons.org/licenses/by-nc-nd/4.0/}\\
$^1${\small Edinburgh Napier University, 10 Colinton Road, Edinburgh, EH10 5DT, United Kingdom.} \\
$^2${\small CNRS, LUTH, Observatoire de Paris-Meudon, 5 Place Janssen, 92190, Meudon, France.}} 

\maketitle
\begin{abstract}
Based on laboratory based growth of plant-like structures from inorganic materials, we present new theory for the emergence of plant structure at a range of scales dictated by levels of ionization, which can be traced directly back to proteins transcribed from genetic code and their interaction with external sources of charge in real plants.  

Beyond a critical percolation threshold, individual charge induced quantum potentials merge to form a  complex, interconnected geometric web, creating macroscopic quantum potentials, which lead to the emergence of macroscopic quantum processes.  The assembly of molecules into larger, ordered structures operates within these charge-induced coherent bosonic fields, acting as a structuring force in competition with exterior potentials.  Within these processes many of the phenomena associated with standard quantum theory are recovered, including quantization, non-dissipation, self-organization, confinement, structuration conditioned by the environment, environmental fluctuations leading to macroscopic quantum decoherence and evolutionary time described by a time dependent Schr\"odinger-like equation, which describes models of bifurcation and duplication.  

The work provides a strong case for the existence of quintessence-like behaviour, with macroscopic quantum potentials and associated forces having their equivalence in standard quantum mechanics.  
The theory offers new insight into evolutionary processes in structural biology, with selection at any point in time, being made from a wide range of spontaneously emerging potential structures (dependent on conditions), which offer advantage for a specific organism.  This is valid for both the emergence of structures from a prebiotic medium and the wide range of different plant structures we see today.  

\end{abstract}

\section*{\centering{1. Introduction}}

The fundamental mechanisms by which molecules assemble into the diverse range of structures that describe living systems remains an open question.  As suggested by Schr\"odinger \cite{Schrodinger1944}, it should be possible to describe all these processes from first principles as pure physical processes.  These initial ideas, along with many other notable contributions, have over the intervening years inspired a substantial body of work in areas such as metabolic networks, non-linear behaviour of complex biological systems and morphogenesis.  A few notable examples, which provide a backdrop to the present paper, include references \cite{Turing1952}-\cite{Kirschner2005}.  However, despite significant progress in these fields, a full understanding of the complexity associated with the genome and its interactions with the environment in dictating structure and function remains illusive, even for relatively simple biological systems where the entire genome has been mapped.

An interesting step in solving this challenge relates to work by Fr\"ohlich \cite{Frohlich1968} who made a connection between long range coherent molecular excitations and long-range biological order.  This work is supported by Prigogine and coworkers \cite{Prigogine1969,Prigogine1973,Prigogine1991,Prigogine1997}, suggesting that `many body' complex systems should be described by models based on probability density, suggesting that complex macroscopic systems are irreducibly probabilistic in a manner equivalent with Quantum Mechanics (QM).  This work leads on to the identification of the importance of collective correlative effects, which contrasts with coarse-grained models, which assume that trajectory dynamics are the whole story, involving an averaging process over the underlying dynamics.  However, whilst this work offers important insights into what follows in the present paper, attempts to build a unified classical/quantum theory have not yet converged toward a formalism which could be practically used.

This general concept of collective correlation emerged again in a recent review \cite{Goushcha2014}, suggesting that a `flux of energy or matter through a system' enables its transition to a new ordered state.  These different ideas sit at the heart of the study of complexity in extended dissipative systems contained within the framework of `self-organized criticality'.  The concept has applicability across a range of non-equilibrium processes from biological systems to plasmas.  However, despite significant progress in this field, a number of key questions around exactly how these processes work in practice remain unresolved \cite{Sharma2015}.  

The focus of the present paper is to address key gaps in our knowledge identified in earlier work by reconsidering the concepts of long-range excitation, self-organized criticality and the fundamental physical principles, which underpin self-organisation.  At the heart of this process we need to explain why we get any structure at all.  According to the second law of thermodynamics, in the absence of a force acting on a system, matter should dissipate, rather than assemble into structures with long-range order.  

We start with the suggestion that a significant first step in the identification of a `missing force' associated with quintessence-like behaviour \cite{Witten2002} driving self-organisation has already been taken \cite{Auffray2008,Nottale2008}.  The fresh insight in this work includes the introduction of the concept of `collective macroscopic quantum forces' playing a potentially critical role in the emergence of biological structure, an idea developed within the theory of scale relativity \cite{Nottale1989}-\cite{Nottale2014b}.  

A key objective of the current work is to explore, test and validate the theoretical principles proposed in \cite{Auffray2008,Nottale2008}. However, whilst this earlier work considered systems biology at a general level, an in depth study of such a large and diverse field would be impractical within the constraints of a single paper.   The work is therefore confined to a limited set of experimental and theoretical developments focussed on the physical principles underpinning the emergence of structure in plants.  At the end of the paper we then consider the potential `universal applicability' of the principles established, not just to plants, but all living systems, as well as considering how the principles driving self-organization could be applied in a new theoretical approach to the development and control of structure in materials development.

Taking a systematic approach, Section 2 summarizes key theoretical concepts developed within the theory of scale relativity, which form the basis for new ideas developed in subsequent sections.  It begins with an introduction to basic principles followed by a description of the geometric foundations of quantum mechanics (QM), which leads onto a derivation of a generalized Schr\"odinger equation, including an equivalent fluid mechanics representation, from first principles.

In section 3 we briefly consider the concept of competing quantum and dissipative systems and the geometric origins of quantum decoherence, whilst Section 4 considers experimental evidence and theory of quantum and dissipative processes to explain the emergence of the diverse array of structures that we observe in plants.

\section*{\centering{2. Theoretical framework}}

\section*{2.1. Basic principles }

The theory of scale relativity shares the basic principles of covariance, the geodesic principle and the principle of equivalence that underpin the theory of general relativity, but re-formulated in a new context to include QM.  At the most fundamental level, the theory challenges the Gauss hypothesis of local flatness, which underlies Riemannian geometry, i.e., the apparently smooth geodesics of space-time at the macro-scale described by general relativity are an incomplete description of the structure of space-time at the micro-scale. 

If we are to understand QM in terms of space-time geometry, we need to rethink its structure in a way that reflects our understanding of the mechanics.  In simple terms, the hypothesis states that the structure of space-time has both a smooth (differentiable) component at the macro-scale and a chaotic, fractal (non-differentiable) component at the micro-scale.  At the macroscale, the fractal component and its influence is small and generally considered unimportant in classical physics.  However, at the microscale, the fractal component and its influence dominate, with quantum laws originating in the underlying fractal geometry of space-time, the space transition taking place at the de Broglie length scale $\lambda_{deB}=\hbar/p$,  whilst the time transition is $\hbar/E$.

To understand the implication of a fractal space-time we need to consider the scale dependence of the reference frames \cite{Celerier2001}.  This means adding resolution $\varepsilon$ to the usual variables (position, orientation, motion) defining the coordinate systems.  However, resolutions can never be defined in an absolute way.  Only their ratio has a physical meaning, allowing an extension of the principle of relativity to that of scales  \cite{Nottale2011,Celerier2001}.

\subsection*{2.2. Geometric foundations}

The transition of a system from the classical to the quantum regime occurs when three critical conditions are satisfied \cite{Nottale2011}.  The first is that the paths or trajectories are infinite in number, leading to a  non-deterministic and probabilistic, fluid like description in which the velocity $v(t)$ on a particular geodesic is replaced by a Eulerian velocity field $v(x,t)$, where the concept of a single trajectory has no meaning. The second, that the paths are fractal curves, transforming the velocity field into a fractal velocity field 
\beq
V=V(x,t,dt).  
\label{fracvelfielddiff}
\eeq
The velocity field is therefore defined as a fractal function, explicitly dependent on resolutions and divergent when the scale interval tends to zero.  The third condition relates to the fundamental breaking of a discrete symmetry implicit in differentiable physics (the reflection invariance on the differential element of [proper] time), which leads to two fractal velocity fields $V_+(x,t,dt)$ and $V_-(x,t,dt)$, which are no longer invariant under transformation ${|dt|}\to-{|dt|}$ in the non-differentiable case.  These velocity fields may in turn be decomposed, i.e.,
\beq
V_+ =v_+(x,t)+w_+(x,t,dt),
\label{eq.V++}
\eeq
\beq
V_- =v_-(x,t)+w_-(x,t,dt).
\label{eq.V--}
\eeq

The (+) and (-) velocity fields comprise a `classical part' $(v_+,v_-)$ which is differentiable and independent of resolution, and fractal fluctuations of zero mean $(w_+, w_-)$, explicitly dependent on the resolution interval $dt$ and divergent at the limit $dt\to0$. 

  A simple and natural way to account for this doubling consists in using complex numbers and the complex product \cite{Nottale2011}.  The three properties of motion in a fractal space lead to a description of a geodesic velocity field in terms of a complex fractal function.  The full complex velocity field reads
\beq
\widetilde{V}={\widehat V} + {\widehat W} = \left(\frac{v_{+}+v_{-}}{2} -i \, \frac{v_{+}-v_{-}}{2} \right) + \left(\frac{w_{+}+w_{-}}{2}-i \, \frac{w_{+}-w_{-}}{2}\right).
\label{FCVF}
\eeq

The jump from a real to a complex description is the origin of the real and imaginary components in the wave function \cite{Nottale2011}.  However, as we show in what follows this is not constrained to the microscale.

\subsection*{2.3. A geodesic approach to quantum mechanics}

The origins of the hypothesis of space-time as a fractal fluid, can be traced back to Feynman \cite{Feynman1965}, who suggested that the typical quantum mechanical paths that are the main contributors to the `path integral', are infinite, non differentiable and fractal (to use current terminology).  This is in agreement with a number of papers for both non relativistic and relativistic quantum mechanics, which have confirmed that the fractal dimension ($D_F$) of the paths is $D_F=2$ \cite{Abbott1981,Kraemmer1974,Campesino1982,Allen1983,Ord1983,Nottale1984}.

If we consider elementary displacements along these geodesics, $dX_\pm=d_\pm x+d\xi_\pm$. In the critical case $D_F=2$, for the geodesics in standard QM, this reads 
\beq
d_{\pm} x= v_{\pm} \; dt, \;\;\;
\eeq
\beq
d\xi_{\pm}=\zeta_{\pm} \, \sqrt{2 \widetilde{D}}  \, |dt|^{1/2}.
\label{eq.20bis}
\eeq
$d\xi$ represents the fractal fluctuations or fractal part of the displacement $dX$. This interpretation corresponds to a Markov-like situation of loss of information from one point to another, without correlation or anti-correlation. Here $\zeta_{\pm}$ represents a purely mathematical dimensionless stochastic variable such that $\langle\zeta_{\pm}\rangle=0$ and $\langle\zeta_{\pm}^2\rangle=1$, the mean $\langle \rangle$ being described by its probability distribution.  $\widetilde{D}$ is a fundamental parameter which characterizes the amplitude of fractal fluctuations.  Since $d\xi$ is a length and $dt$ a time, it is given by the relation
\beq
{\widetilde D}= \frac{1}{2}\, \frac{\langle\enspace \!\!d \xi^2 \!\!\enspace\rangle}{dt},
\label{diffusionrelation}
\eeq

its dimensionality is therefore $[L^2T^{-1}]$. 

When considering the geodesics of a fractal space, the real and imaginary parts of the velocity field can be expressed in this case in terms of the complex velocity field\footnote{For simplicity ${\widehat W}$ is not considered here since it vanishes when taking the mean \cite{Nottale2011}.}.
\beq
{\widehat V}= V- i U.
\label{compvelfield}
\eeq
This equation captures the essence of the principle of relativity in which any motion, however complicated and intricate the path, should disappear in the proper reference system
\beq
{\widehat V}= 0.
\label{eq.V0}
\eeq
We now introduce the complex `covariant' derivative operator $\widehat{d}/dt$, which includes the terms which allow us to recover differentiable time reversibility in terms of the new complex process \cite{Nottale1993,Nottale2011} 
\beq
\frac{\widehat{d}}{dt} = \frac{1}{2} \left(\frac{d_+}{dt}+ \frac{d_-}{dt}\right)-\frac{i}{2}\left(\frac{d_+}{dt} - \frac{d_-}{dt}\right).
\label{eq.complexprocess}
\eeq
Applying this operator to the position vector yields the differentiable part of the complex velocity field 
\beq
{\widehat V} = \frac{\widehat{d}}{dt} \,x = V -i U = \frac{v_+ + v_-}{2} - i 
\;\frac{v_+ - v_-}{2}.
\label{complex vF}
\eeq
Deriving Eq.~(\ref{eq.V0}) with respect to time, it takes the form of a free strongly covariant geodesic equation.
\beq
\frac{\widehat{d}}{dt} \,{\widehat V} =0.
\label{eq.37bis}
\eeq
In the case of a fractal space, the various effects can be combined in the form of a complex covariant derivative operator \cite{Nottale1993,Nottale2011},

\beq
\frac{\dfr}{dt} = \frac{\partial}{\partial t} + {\widehat V}. \nabla - 
i {\Ds} \Delta \; ,
\label{comptimderiv}
\eeq

which is analogous with the covariant derivative $D_jA^k =  \partial_j A^k + \Gamma_{jl}^k A^l$ replacing $\partial_jA^k$ in Einstein's general relativity (in the sense that it allows one to implement the principle of covariance).  This allows us to write the fundamental equations of physics under the same form they had in the differentiable case \cite{Nottale2011}, i.e. the fundamental equation of dynamics becomes, 
\beq
m \, \frac{\widehat{d}}{dt} \,{\widehat V} =- \nabla \phi,
\label{eq.dyn}
\eeq
which is now written in terms of complex variables and of the complex time derivative operator.  Which, in the absence of an exterior field $\phi$, is a geodesic equation.

In describing the three essential conditions required for quantum behaviour, we have shown how the first two conditions are shared with diffusive systems, typical of Brownian motion and more generally of Markov processes, with Eq.~(\ref{eq.20bis}) describing fluctuations in Brownian motion.  In these types of processes, a particle follows a random walk in which both direction and distance are uniformly distributed random variables.  In moving from a given position in space to any other, the path taken by the particle has a very high probability to fill the whole space before reaching its destination, hence as with QM, Brownian motion is also characterized by $D_F=2$.  However, whilst the two processes share the first two conditions, the third condition - the complex velocity field differentiates between quantum and dissipative systems, an issue that we revisit in Section 3.

\subsection*{2.4. The Schr\"odinger equation and its equivalent fluid representation with a quantum potential.}

After expansion of the covariant derivative, the free-form motion equations of general relativity can be transformed into a Newtonian equation in which a generalized force emerges, of which the Newton gravitational force is an approximation.  In an analogous way, the covariance induced by scale effects leads to a transformation of the equation of motion, which, as we demonstrate through a number of steps in Section 7.1 (supplementary material), leads to a generalized Schr\"odinger equation  Eq.~(\ref{eq.Schrodinger-diff}).

\beq
{\widetilde D}^2 \Delta \psi + i {\widetilde D} \frac{\partial}{\partial t} \psi - 
\frac{\phi}{2m}\psi = 0,
\label{eq.Schrodinger-diff}
\eeq
where $\widetilde D$ identifies with the amplitude of the quantum force, which is more general than its standard QM equivalent $(\hbar/2m)$, accommodating both the one body and many body case (either distinguishable or indistinguishable particles) \cite{Nottale2009,Nottale2011}, as well as the possibility of macroscopic values.

This equation (\ref{eq.Schrodinger-diff}) can alternatively be written as a combination of Euler and continuity equations (Eq's.~\ref{Euler1} \ref{continuity1}).

\beq
m \l( \frac{\d}{\d t} + V. \nabla\r) V=- \nabla \phi +2 m \Ds^2\nabla \l( \frac{\Delta\sqrt{P}}{\sqrt{P}}\r),
\label{Euler1}
\eeq  

\beq
\frac{\d P}{\d t} + \text{div} ( P V)=0,
\label{continuity1}
\eeq

where Eq.~(\ref{Euler1}) describes a fluid subjected to an additional quantum potential $Q$ Eq.~(\ref{Eulerquantpot1}). 
\beq
Q=-2 m \Ds^2 \: \frac{ \Delta \sqrt{P}}{\sqrt{P}}.
\label{Eulerquantpot1}
\eeq

This system of equations, the detailed derivation of which is outlined in Section 7.2 (supplementary material), is equivalent to the classical system of equations of fluid mechanics (with no pressure and no vorticity), except for the change from a density of matter to a density of probability.  The potential energy term $Q$, is a manifestation of the fractal geometry and probability density, with fractal space-time fluctuations (at the micro-scale) leading to the emergence of a `fractal field', a potential energy (`quantum potential') and a quantum force which are directly analogous with the geometric origins of the gravitational field, gravitational potential and gravitational force which emerge as a manifestation of the curved geometry of space-time.

The potential energy $Q$ is a generalization of the quantum potential in standard QM, which relies on  $\hbar$, but which is here established from the geodesic equation as a fundamental manifestation of the fractal geometry.  It is implicitly contained in the Schr\"odinger form of the equations, but only explicit when reverting to the Euler representation.  A particle is therefore described by a wave function (constructed from the geodesics), of which only the square of the modulus $P$ is observable, the `field' being given by a function of $P$, or density of matter $\rho$ (since $\rho = PM$, i.e., $\psi = \sqrt{ \rho} \,\times {e}^{iA/\hbar}$ when M =1).

This new geometric approach to quantum theory offers some important new insights.  Classical quantities relate to the quantum world as averages (Ehrenfest theorem).  Conversely quantum properties remain at the heart of the classical world.  We have shown that the action of fractality and irreversibility on small time scales can manifest itself through the emergence of a macroscopic quantum-type potential energy, in addition to the standard classical energy balance.  This potential energy leads to the possibility of `new'  macroscopic quantum effects, no longer based on the microscopic constant $\hbar$, which are normally masked by classical motion, but can be observed given the right conditions, a subject we address in detail in Section 4.

\subsection*{\centering{3. Quantum decoherence}}

\subsection*{{3.1. The transition from diffusion to quantum coherence}}

In Section 2.2 we differentiate between a diffusive system described by a fractal velocity field Eq.\ref{fracvelfielddiff} and a quantum system described by a complex velocity field Eq.(\ref{FCVF}).  We now revisit these two systems and their role in quantum decoherence.  

In section 7.3 (supplementary material) we consider the details which lead to the description of a diffusion system in the form of a Euler equation Eq.(\ref{DP1})
\beq
\l( \frac{\d}{\d t} + V. \nabla\r)V=- 2 D^2 \, \nabla \l( \frac{\Delta \sqrt{P}}{\sqrt{P}}\r).
\label{DP1}
\eeq
This compares in a striking way with the quantum equivalent in the free case Eq.~(\ref{eq.621}) \cite{Nottale2008}. 

\begin{equation}
\l(\frac{\partial}{\partial t} + V.\nabla\r) V  = +2{\Ds}^2 \, \nabla \l(\frac{\Delta \sqrt{P}}{\sqrt{P}}\r).
\label{eq.621}
\end{equation}
The two equations demonstrate a clear equivalence between a standard fluid subjected to a force field and a diffusion process with the force expressed in terms of the probability density at each point and instant.  

The `diffusion force' derives from an external potential
\beq
\phi_{\rm diff}=+2D^2 {\Delta \sqrt{P}}/{\sqrt{P}}.
\label{diffpot}
\eeq			
which introduces a square root of probability in the description of a totally classical diffusion process.  The quantum force is the exact opposite, derived from the 'quantum potential', which is internally generated by the fractal geodesics 
\beq
Q/m=-2{\Ds}^2 {\Delta \sqrt{P}}/{\sqrt{P}}.
\label{quantpot}
\eeq

This interpretation offers new insights into quantum decoherence in both standard QM and in macroscopic quantum systems such as High Temperature Super Conductivity (HTSC) \cite{Turner2015}, where the two forms of potential energy exist and compete in quantum systems, summarized by the total system-environment Hamiltonians ($H_S$-$H_E$), and their interaction ($H_{int}$)

\beq
H=H_S+H_E+H_{int}.
\label{decoherence}
\eeq

The description of competing quantum and dissipative forces fits well with decoherence theory described by the model of `quantum Brownian motion' \cite{Zurek2003,Schlosshauer2014}. During the decoherence process, the time evolution of position-space and momentum-space is reflected in the superposition's of two Gaussian wave packets \cite{Schlosshauer2014}. Interference between the two wave packets is represented by oscillations between the direct peaks. Interaction with the environment damps these oscillations. While the momentum coordinate is not directly monitored by the environment, the intrinsic dynamics, through their creation of spatial superposition's from superposition's of momentum, result in decoherence in momentum space as the two valuedness associated with the complex component $iU$ of the velocity field (Eq.\ref{compvelfield}) begins to break down.  This leads to the emergence of pointer states, which are minimum-uncertainty Gaussians (coherent states), well-localized in both position and momentum, thus approximating classical points in phase space \cite{Zurek2003,Joos2003,Kubler1973,Paz1993,Zurek1993,Diosi2000}.  

This process appears to have been well described.  However, an important question remains relating to the origin of `pointer states'  in the decoherence process \cite{Zurek2003,Schlosshauer2014,Zurek2013}.  Within the context of a fractal fluid of geodesics constituting a standard or macroscopic quantum system $\rho=|\psi|^2$, we set the hypothesis that the emergence of pointer states is linked to a fundamental root structure\footnote{A ubiquitous characteristic of fractal networks.} underpinning the fractal velocity field $\widehat V$.  During the process of decoherence, the fractal velocity field collapses to its more stable roots.  These roots form the preferred set of states of an open system most robust against environmental interaction, accounting for the transition from a probabilistic to a deterministic classical description.  A full description of this concept within the context of standard QM, falls outside of the scope of the current paper.  However, we consider this issue here, within the context of macroscopic quantum systems.

\subsection*{{3.2. Macroscopic quantum decoherence}}

Fractal stochastic fluctuations, dominate the quantum realm.  However they remain at the heart of the classical world as zero averages, the fractal component being masked by the classical contribution.  In the scale relativistic foundation of standard quantum mechanics, the relation $d\xi =\eta η \sqrt 2\Ds|dt|$ is considered valid at all scales \cite{Nottale2011}. The de Broglie transition to the classical realm is but an effective transition, which comes from the domination of the classical term $dx = vdt$ over the fractal term $d\xi$ at scales $\delta x >\hbar/mv$. The fractal part remains at macroscopic scales, masked by classical motion. 

The fractal-nonfractal transition (the boundary between quantum coherence and classic systems) is blurred by the numerous cases of mesoscopic interference experiments and macroscopic quantum phenomena such as  conventional super conductivity (SC).  The blurring of the transition is partly due to the fact that the relative contributions of classical and fractal components depends on the value of the transition, which is itself relative, depending in particular on the state of motion of the reference system. Namely, the de Broglie length-scale is $ \lambda_{\rm deB}=\hbar/p=\hbar/mv$ for a free particle, while the de Broglie thermal scale is $\lambda_{\rm th}=\hbar/(2m kT)^{1/2}=\hbar/(m\langle v^2 \rangle^{1/2})$, and the de Broglie time $\tau_{\rm deB}=\hbar/E=\hbar/\frac{1}{2}m v^2$).  By way of illustration we consider the $p$ type cuprates which represent the most widely studied case of high temperature super conductors \cite{Turner2015}.  As with most cases of conventional super conductors (SC), we are not dealing with a fully coherent system since the medium (an antiferromagnetic cuprate structure) remains classical.  In this instance, electron-pair (e-pair) coupling energies are significantly higher than conventional SC, leading to more thermally stable, but localized e-pairs.  Below a critical temperature $T_c$, the transition from localization to coherence is facilitated by macroscopic quantum potentials (MQP's) determined by a macroscopic de Broglie scale $\lambda_{deB}=2\Ds/v$), dependent on \cite{Turner2015}:

\begin{itemize}
 \item a scale free distribution of dopants (charges), the frequency and extent of fluctuations being dependent on the $D_F$ and correlation length of the scale free network.   

\item the pressure term, which is a function of matter density $\rho$, so the velocity field $V$ of a fluid in potential motion is described in terms of a complex function $\psi=\sqrt{\rho} {\times e^{iA/\hbar}}$. 
\end{itemize}

To summarize, macroscopic quantum coherence is only linked to bosons \cite{Turner2015}.  In Section 4 we explore the role of coherence and decoherence of bosonic fields in the determination of plant structures at different scales.

\subsection*{\centering{4. Macroscopic quantum mechanics:  the origins of self-organization}}

\subsection*{4.1. Background}

A Schr\"odinger-type equation is characterized by the existence of stationary solutions, yielding well-defined peaks of probability linked to quantization laws, which are themselves a consequence of the limiting (or environmental) conditions, of the forces applied and of the symmetries of the system.  Peaks of probability density distribution are interpreted as a tendency for a system to make structures, allowing prediction of the most probable structures among an infinity of close possibilities.  This process is analogous with living systems, where morphologies are acquired through growth processes, which can be described in terms of an infinite family of virtual, fractal and locally irreversible, fluid-like trajectories, suggesting the possibility that they can be written in the form of a macroscopic Schr\"odinger equation (Eq.\ref{eq.Schrodinger-diff}), leading to the emergence of quantized structures \cite{Nottale2008,Nottale2011}.   

If the hypothesis is correct, once these processes are better understood, it should be possible to apply this knowledge to grow flowers and other plant like structures in non biological systems from a range of bio-polymers and/or inorganic matter.  During the process of developing such an approach, the growth of a range of $BaCO_3-SiO_2$ based plant-like structures from solutions of $BaCl_2$ and $Na_2SiO_3$ was reported \cite{Norrduin2013}.  The work demonstrated convincingly that alongside a range of boundary conditions, levels of atmospheric $CO_2$ played a key role in structure formation.  However, whilst there was speculation that `chemical fields' play a role in the process, an explanation of the mechanism by which ordered structures (rather than e.g., crystalline or disordered systems) emerged was not fully elucidated.  

We set the hypothesis that these plant-like structures are in fact driven by the equivalent of a macroscopic quantum mechanics-type process.  However, to validate this interpretation, theory dictates that the internal mesoscale geometry of observed structures must be predominantly fractal if they are to meet criteria for the emergence of a macroscopic quantum potential (Eq.\ref{Eulerquantpot}), a prerequisite for macroscopic quantum processes.

Unfortunately the resolution of images published by Norrduin {\it{et al}} \cite{Norrduin2013} was not at a level where a fractal mesoscale structure could be confirmed.  To validate, or refute our hypothesis on fractality, key aspects of the work were replicated and the emergent structures analysed using high resolution Field Emission Scanning Electron Microscopy (FE-SEM).  A successful outcome would clarify our understanding of the mechanisms that dictate the emergence of different structures, which may in turn lead to better insights into plant morphogenesis.

\subsection*{4.2. Experimental work}

\subsection*{4.2.1. Method}

Stage 1. Growth of structures was conducted in a 250 ml glass beaker, following experimental conditions described by Norrduin {\it{et al}} \cite{Norrduin2013}, with aqueous solutions of $BaCl_2$ (19.1mM) and $Na_2SiO_3$ (8.2 mM) prepared at pH 11.2.  Growth was carried out for one hour at room temperature ($\approx 18^0$C) on polished aluminium plates of the same size as standard microscope slides.  The plates were stood at an angle, with the solution covering $\approx 30\%$ of its height, in a system open to atmospheric $CO_2$.

Stage 2.  The impact of $CO_2$ levels on growth was tested to its limits by repeating the work in a closed system, to allow control over levels of $CO_2$ during the growth period.  In the first case, ambient $CO_2$ concentration was minimized by continuous purging of the system (the atmosphere above the solution) with nitrogen gas.  In a second trial, the system was purged with a continuous flow of gaseous $CO_2$.

Stage 3.  Stage 1 was repeated with ambient concentrations of $CO_2$ in a fridge at $4^o$C.  This particular condition varied from the method described by Norrduin {\it{et al}} \cite{Norrduin2013}, where only the solution was cooled to $4^o$C .

After each growth stage, emergent structures were carefully air dried at room temperature following the description by Norrduin {\it{et al}} \cite{Norrduin2013}, before FE-SEM analysis.

\subsection*{4.2.2. Results and discussion}

\begin{figure}[!ht]
\begin{center}
\includegraphics[width=16cm]{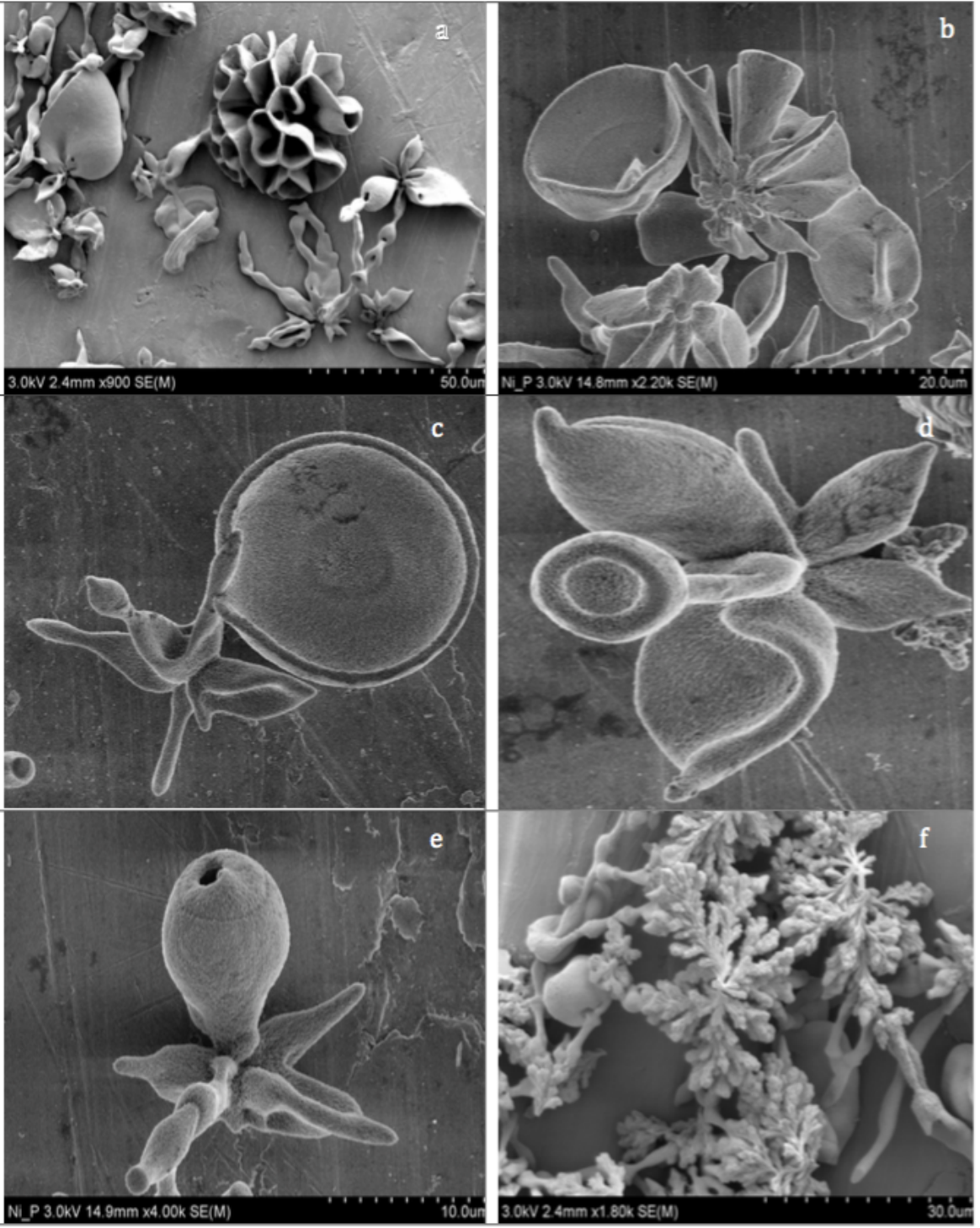} %
\caption{\small{$$ Observed plant-like structures in ambient $CO_2$.}}
\label{Plant structures}
\end{center}
\end{figure}

As reported previously by Norrduin {\it{et al}} \cite{Norrduin2013}, under Stage 1 conditions, a range of different plant-like structures were observed over the growth region on the plate, from which a sub-sample is reported here.  Fig~\ref{Plant structures}a (900 x magnification) shows examples of stem, leaf, pod and coral-like structures, whilst Fig~\ref{Plant structures}b (2,200 x magnification) illustrates a combination of cone, bowl and leaf-like structures plus more `open', flower-like structures (with discrete petals) growing on top of more ordered structures.  

In Fig~\ref{Plant structures}c ($\approx$ 4000 x magnification), we see a symmetric water-lily-leaf structure, whilst Fig~\ref{Plant structures}d shows a smaller diameter, partially `closed' hemispherical structure at the stem apex.  This process of closure is almost complete in Fig~\ref{Plant structures}e where we see a pod or fruit like structure.

In general, the relative proportions of the mix of different structures varied, depending on proximity to the solution in which the plate was standing.  Objects closest to the solution interface reflected a higher proportion of ordered structures typified by Fig's~\ref{Plant structures}a-e.  However, as distance from the solution increased, a greater proportion of less ordered, fractal structures typified by Fig~\ref{Plant structures}f (1,800 x magnification) emerged. 

\begin{figure}[!ht]
\begin{center}
\includegraphics[width=16cm]{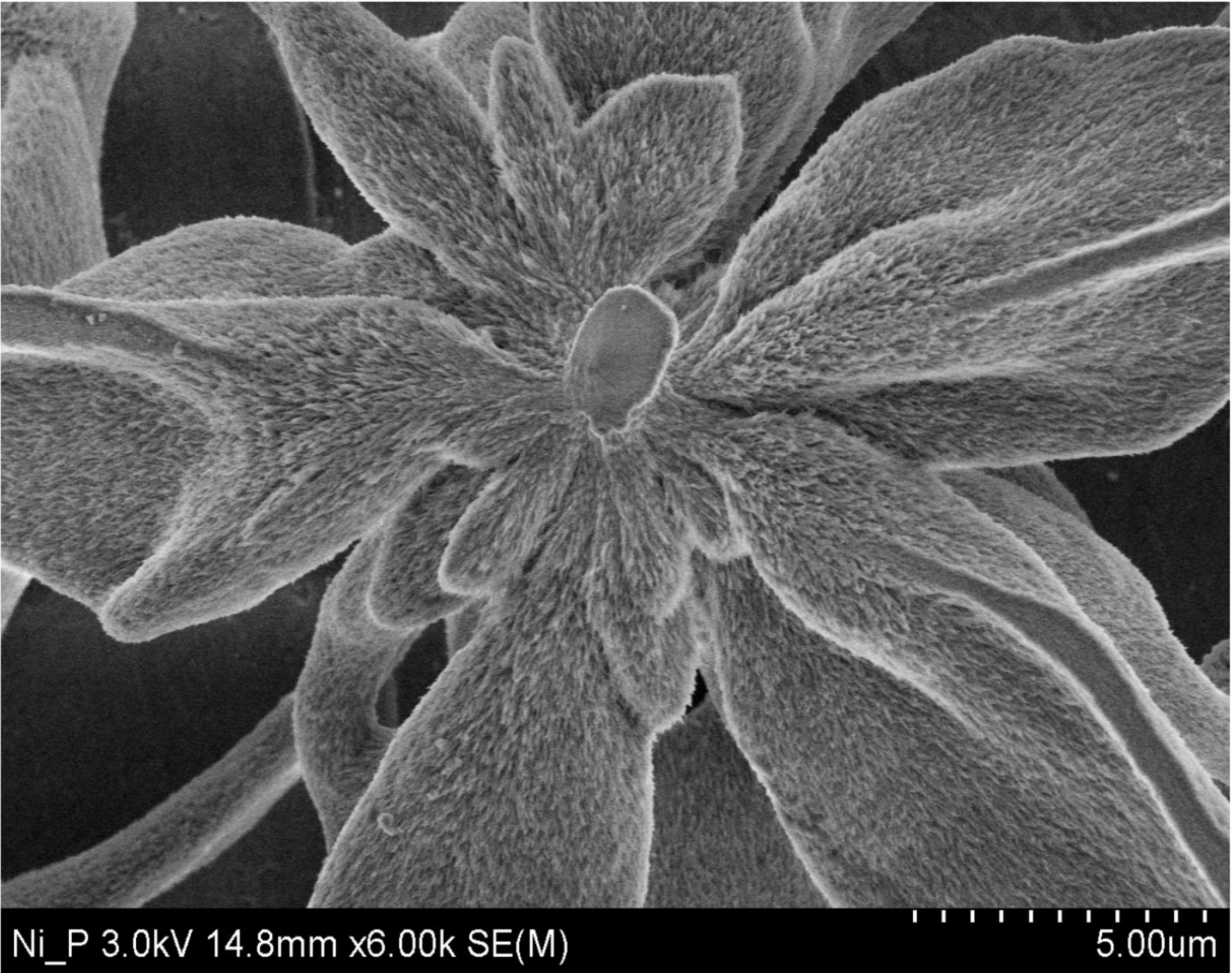} %
\caption{\small{$$ Fractal mesoscale architecture of a flower-like structure}}
\label{fractalflower}
\end{center}
\end{figure}

\begin{figure}[!ht]
\begin{center}
\includegraphics[width=16cm]{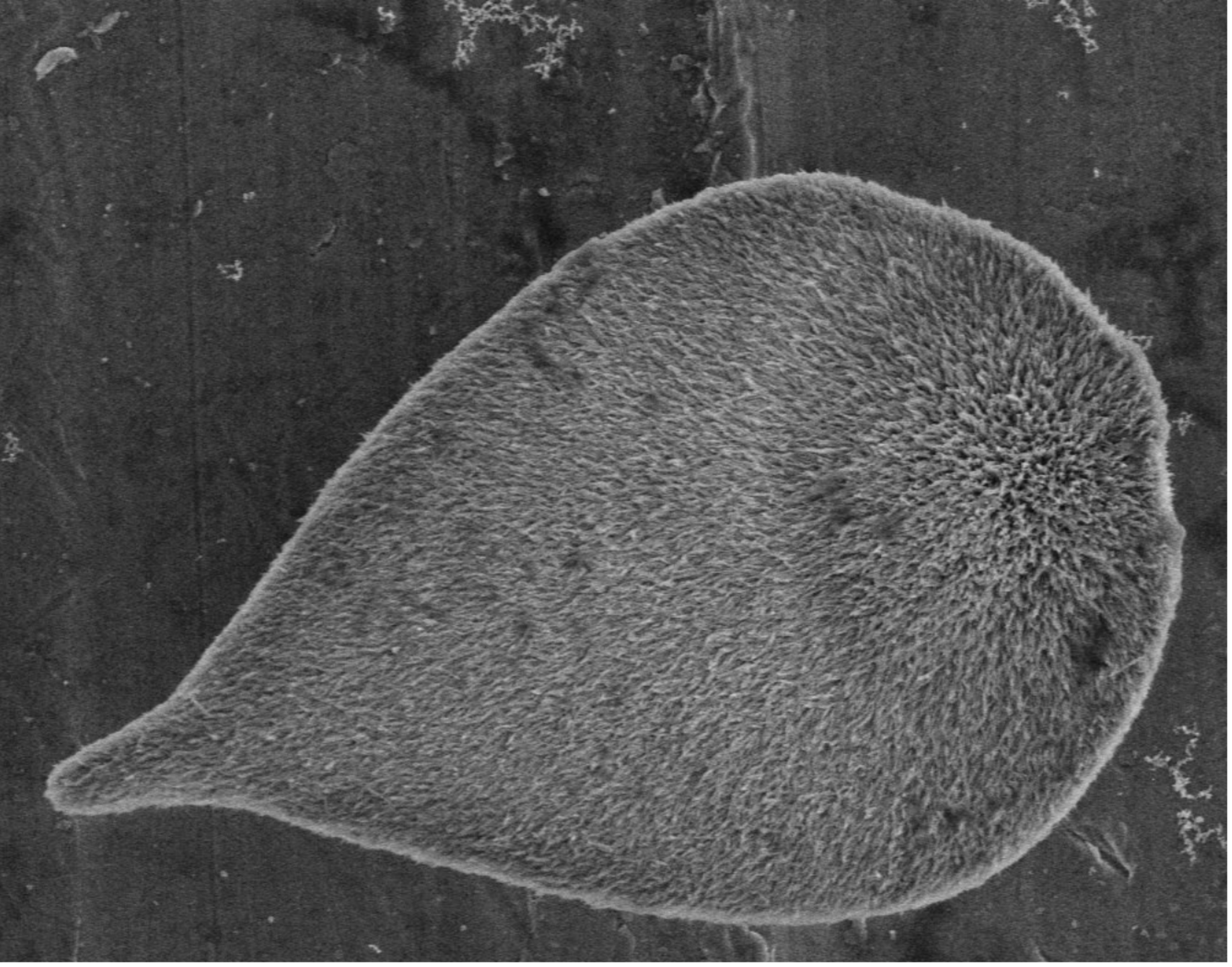} %
\caption{\small{$$ Fractal mesoscale architecture of a leaf-like structure}}
\label{Leaf}
\end{center}
\end{figure}

\begin{figure}[!ht]
\begin{center}
\includegraphics[width=16cm]{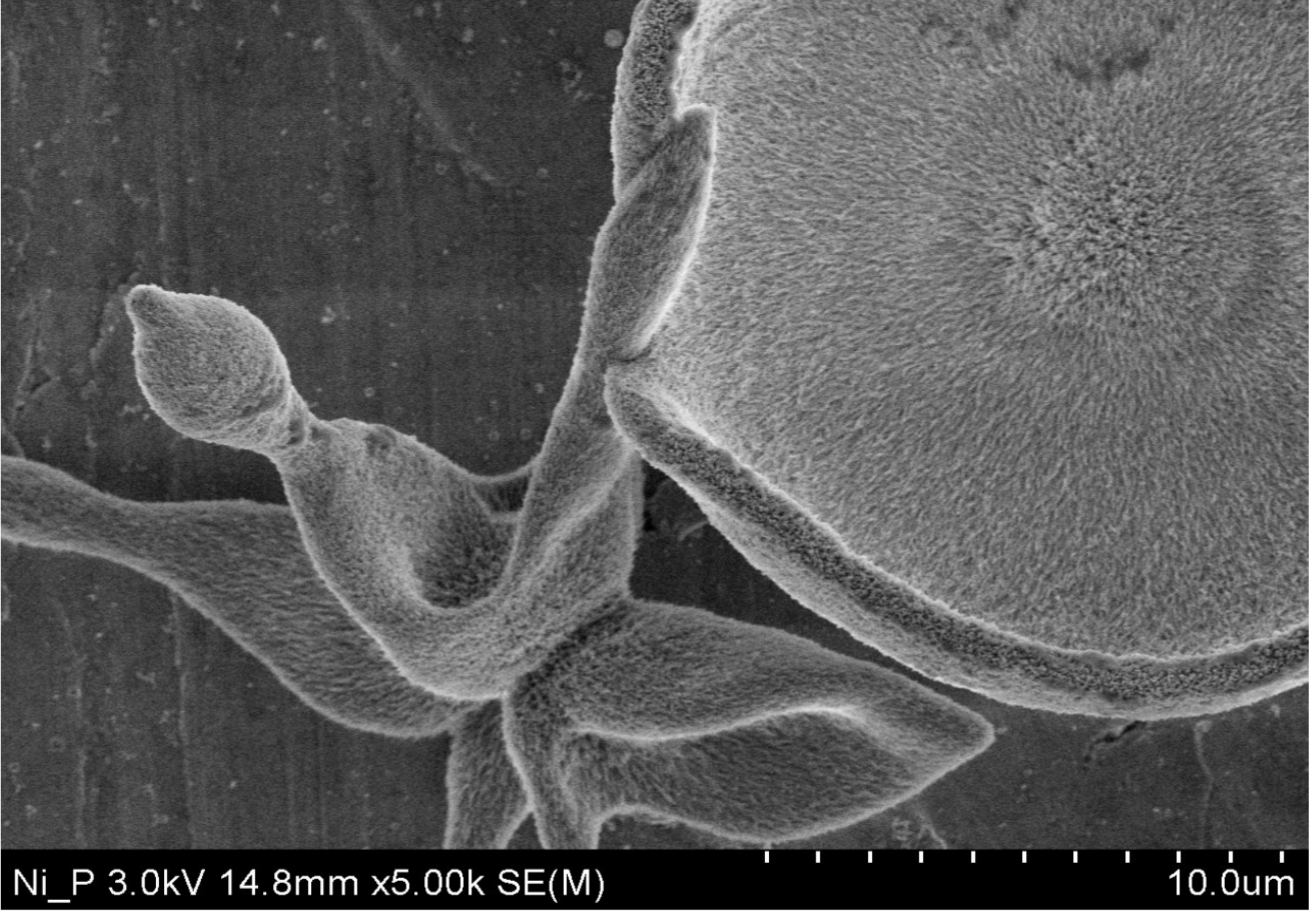} %
\caption{\small{$$ Fractal mesoscale architecture of a Lilly leaf-like structure}}
\label{fractallily}
\end{center}
\end{figure}

\begin{figure}[!ht]
\begin{center}
\includegraphics[width=16cm]{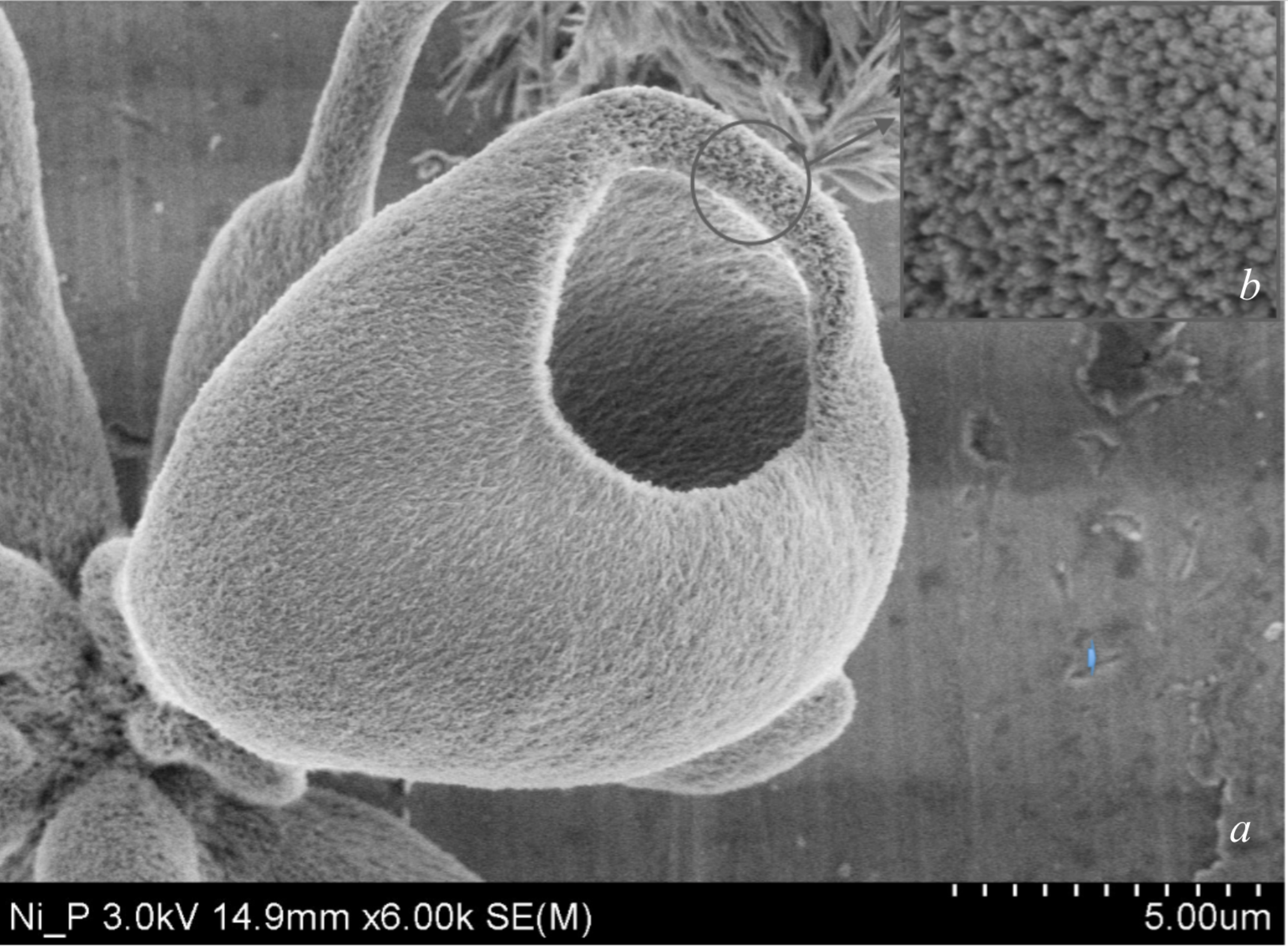} %
\caption{\small{ Fractal mesoscale architecture of a pod (a). Insert (b) shows an enlarged $\approx 1 \mu m^2$ section showing the internal fractal wall structure }}
\label{pod}
\end{center}
\end{figure}

To test our hypothesis that the internal mesoscale structure of these plant-like objects should be fractal we increased image resolution.  Fig~\ref{fractalflower} illustrates an example of a 6000 x magnification of a flower-like structure with discrete petals observed in Fig~\ref{Plant structures}b.  At this resolution we see a distribution of $\approx 10 nm$ diameter fibrils, which we would typically expect from a fractal structure, generally oriented in the direction of growth from a central point.  The same structure was found in the leaf like structures illustrated in Fig~\ref{Leaf} and Fig~\ref{fractallily} (an enlargement of Fig~\ref{Plant structures}c) and in an enlargement of a pod structure (Fig~\ref{pod}a)\footnote{We note that in the structures reported here, acid treatment described by Norrduin {\it{et al}} \cite{Norrduin2013} was not used.  However, some preliminary work was carried out with acid treatment to determine its impact on the mesoscale structures observed.  No change in the fractal mesoscale fractal structure was observed.}, which resembles the structure of the Barrel sponge ({\it{Xestospongia testudinaria}}).  In Fig's~\ref{fractallily} and \ref{pod}a, the rim of the leaf and pod structures indicates an $\approx 1\mu m$ thick wall, with an internal fractal architecture, which is revealed in more detail in Fig~\ref{pod}b.

The results confirm our hypothesis of an internal mesoscale fractal structure in the plant-like structures reported by Norrduin {\it{et al}} \cite{Norrduin2013}, which satisfies a key condition for the emergence of macroscopic order driven by macroscopic quantum processes. 

In describing a mechanism for the emergence of observed structures we first consider the molecular to $nm$ scale of assembly.    Quantum vacuum fluctuations (viewed as a sea of harmonic oscillators),  thermal fluctuations and associated phonons, are inextricably linked and correlated.  Acting collectively as `environmental fluctuations', they have a significant impact on the trajectory and dynamics of unconstrained particles as they interact to form larger structures.  To understand their role in more detail, consider the molecular scale environment created by $CO_2$, identified by Norrduin {\it{et al}} \cite{Norrduin2013} as a critical variable.  

$CO_2$ solvation leads to the release of protons and the subsequent ionization of $BaCO_3-SiO_2$ molecules, with charge density $\rho$ determined by $CO_2$ concentration and temperature $T$, which determines $CO_2$ solubility (increasing $T$ $\rightarrow$ increasing $\rho$).

Repulsive forces between adjacent charged particles influences the dynamics of molecular assembly during growth of the structures observed.  At a simplistic level, increasing $\rho$ leads to an increase in the degree of molecular freedom to interact with environmental fluctuations.  

At an individual level, a single charge is repulsive.  However, as levels of charge increase, charges will cluster loosely, with $\mathring{A}$-scale holes within clusters creating attractive potential well's, which may be interconnected, via channels between them, induced by charge distribution (see Turner and Nottale \cite{Turner2015}, Fig's 3 and 4).

At a local level, clusters of charges constitute a quantum fluid 

\beq
\psi_n=\displaystyle\sum\limits_{n=1}^N\psi_{d_n} 
\label{psin}
\eeq

which is expected to be the solution of a Schr\"odinger equation 
\beq
\frac{\hbar^2}{2m} \Delta \psi_n + i \hbar \frac{ \d \psi_n}{\d t}= \phi\;  \psi_n,
\label{Schrodingercharge}
\eeq 
where $\phi$ is an exterior potential

We now introduce explicitly the probability density (charge density) $\rho$ and phase, which we define as a dimensioned action $A$ of the wave function  $\psi_n= \sqrt{\rho_n} \times e^{{i A_n}/\hbar}$. The velocity field of the   quantum fluid ($n$) is given by $V_n=(\hbar/m) \nabla{ A_n/\hbar}$.  

Following Eq's.(\ref{continuity}) and (\ref{Euler}), we write the imaginary part of Eq.(\ref{Schrodingercharge}) as a continuity equation and the derivative of its real part as a Euler equation.

\beq
\frac{\d V_n}{\d t} + V_n. \nabla V_n =- \frac{\nabla \phi}{m} - \frac{\nabla Q_n}{m},
\label{eq.vsEuler}
\eeq
\beq
\frac{\d \rho_n }{ \d t} + \text{div}(\rho_n V_n)=0.
\label{eq.vscont}
\eeq

where $Q_n$ represents localized quantum potentials, dependent on local fluctuations of the density $\rho_n$ of static charges,
\beq
 Q_n= -\frac{\hbar^2}{2m} \frac{\Delta \sqrt{\rho_n}}{ \sqrt{\rho_n}}.
\label{eq.qd}
\eeq

\begin{figure}[!ht]
\begin{center}
\includegraphics[width=10cm]{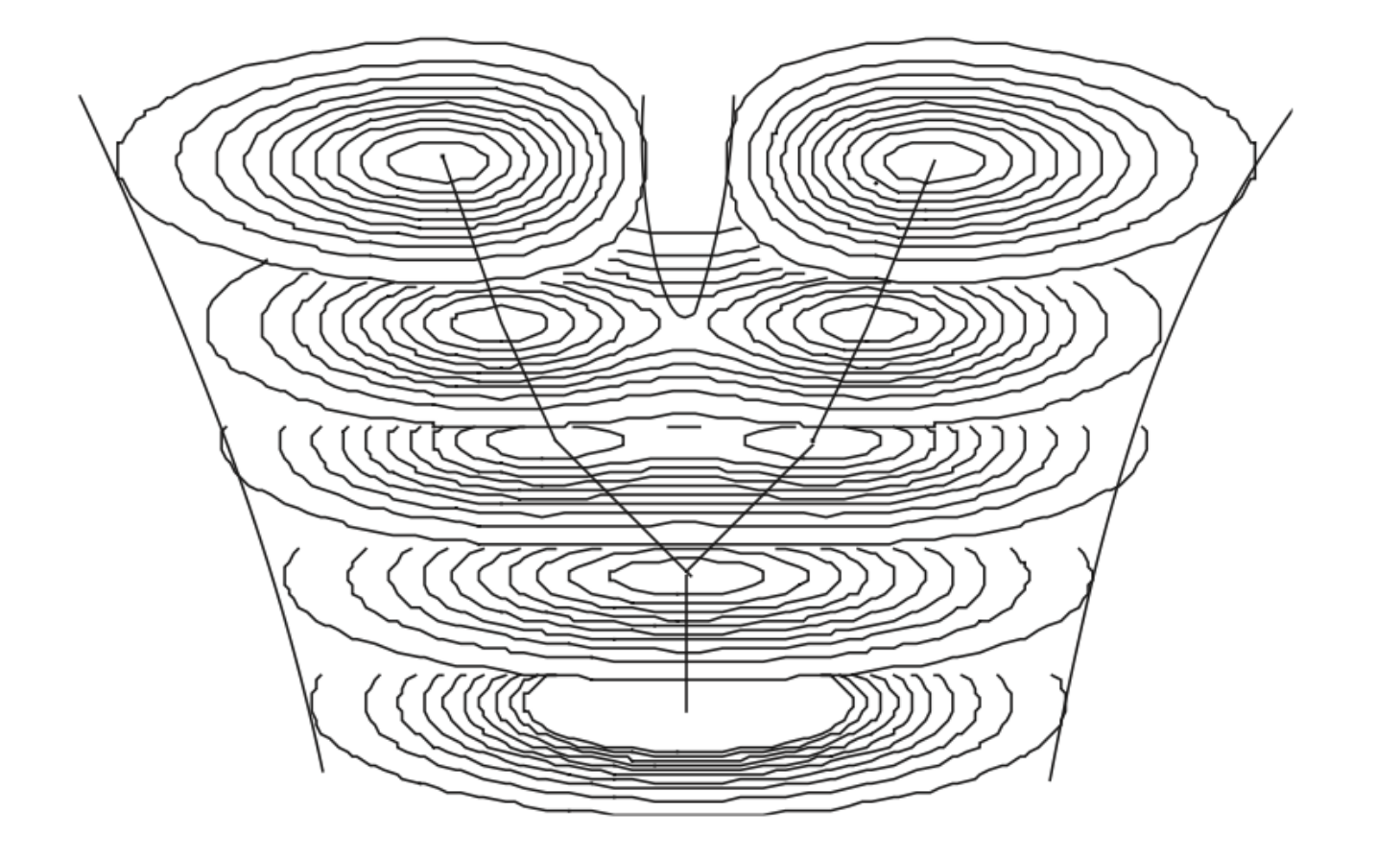} %
\caption{\small{Model of branching/bifurcation, described by successive solutions of the time-dependent 2D Schr\"odinger equation in an harmonic oscillator potential plotted as isodensities} \cite{Nottale2008}.}
\label{Bifurcation.pdf}
\end{center}
\end{figure}

\begin{figure}[!ht]
\begin{center}
\includegraphics[width=12cm]{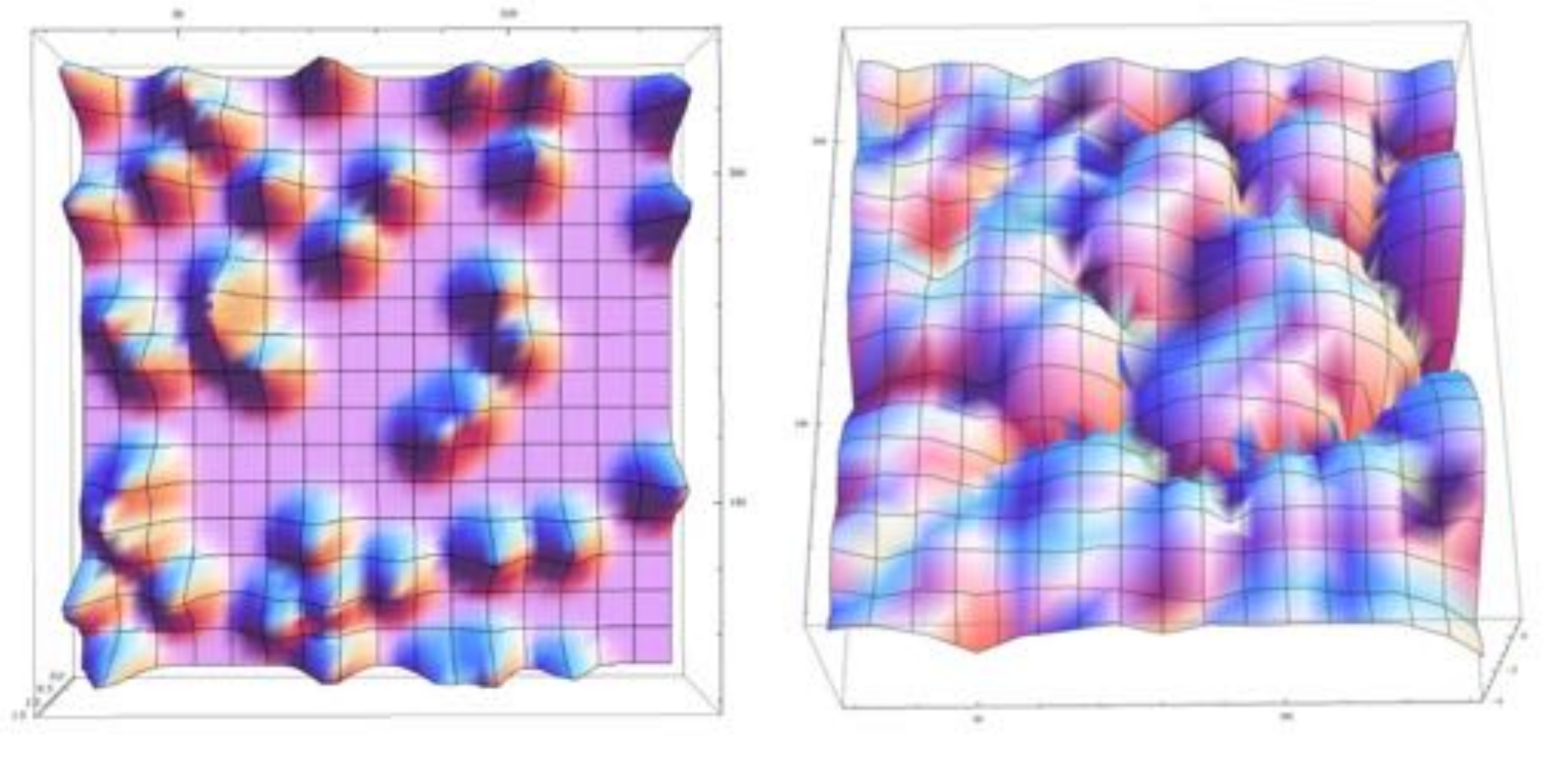} %
\caption{\small{Individual charges collectively create a geometric network of hills and valleys.}}
\label{charges}
\end{center}
\end{figure}

As described in previous work \cite{Nottale2008}, successive solutions during time evolution of the time-dependent Schr\"odinger equation in a 2D harmonic oscillator potential (plotted as isodensities), leads to a model of branching/bifurcation.  This original work was based on a macroscopic Schr\"odinger equation using the macroscopic constant $\Ds$.  However,  it is equally applicable in the standard QM case based on $\hbar$, which we consider here.  Adopting this approach, the energy level varies from the fundamental level ($n=0$) to the first excited level ($n=1$).  As a consequence the system jumps from a one-body to a two-body branched structure (Fig~\ref{Bifurcation.pdf}) which (given sufficient charge), leads to a branched molecular assembly and the emergence of fractal architectures\footnote{We note that whilst the model can be illustrated by an harmonic oscillator potential (2D or 3D) and by box solutions \cite{Nottale2011}, it is a very general feature of solutions of the Schr\"odinger equation, whose fundamental (`vacuum') states show a unique global structure while first excited states generally show a two-body structure.}. During this process, charge induced potential well's will interconnect, creating a fractal network of channels, with the charges acting as roots of the fractal web, which we illustrate in a simplistic two dimensional model (Fig~\ref{charges}).

Extending the scenario in Fig~\ref{charges} to a 3D fractal architecture will lead to the emergence of a fractal distribution of static charges ($\psi_{d}$) and a charge induced fractal velocity field $\widehat V$.   At a critical threshold, destructive interference effects induced by collective charges $\psi_n$, cancel out of most frequencies, leaving a coherent resonant frequency, dictated by the geometry of the fractal network.  Evidence for this type of behaviour in fractal networks is suggested by the emergence of coherent e-pairs in HTSC or photons in Coherent Random Lasing (CRL) \cite{Turner2015}.  The result is an infinite connected web of coherent charge fluctuations defined by $\Ds$, rather than $\hbar$; the extent of fluctuations (correlation length) is determined by the scale of the structure formed.

This has the effect of transforming quanta of fractal fluctuations $\psi_n = \sqrt{\rho_n} \times e^{{i A_n}/\hbar}$ (where $A_n$ is a microscopic action), into macroscopic fluctuations creating the equivalent of a path integral, represented by a macroscopic wave function $\psi_N$

\beq
\displaystyle\sum\limits_{n=1}^N\psi_{d_n} \to  \psi_{N} = \sqrt{\rho_{N}}\times e^{i A_N/2\Ds},
\label{eq.68}
\eeq
where $A_N$ is a macroscopic action and (since $\rho=Pm$) $m=1$, and $Q_N$ is its associated MQP,

\beq
Q_{{N}}  = -2{\Ds}^2 \, \frac{\Delta \sqrt{\rho_{N}}}{\sqrt{\rho_{N}}}.
\label{eq.QMG}
\eeq

We can now re-write the Euler and continuity equations (Eq's \ref{eq.vsEuler} and \ref{eq.vscont})

\beq
\frac{\d V_{N}}{\d t} + V_{N}. \nabla V_{N} =- \frac{\nabla \phi}{m} -\frac{\nabla Q_N}{m},
\label{eq.CMGEuler}
\eeq
\beq
\frac{\d \rho_N }{ \d t} + \text{div}(\rho_N V_N)=0,
\label{eq.CMGcont}
\eeq

which can be re-integrated under the form of a macroscopic Schr\"odinger equation,

\beq
{\Ds}^2 \Delta \psi_N + i {\Ds} \frac{\partial\psi_N}{\partial t} - \l(\frac{\phi}{2}\r)\psi_N = 0
\label{eq.SchrodingerC}.
\eeq

In principle the molecular scale fractal architectures could be expected to grow in a continuous process, until a point where growth is constrained by a natural symmetry breaking at scales $>40 \mu m$, when gravitational forces exceed van der Waals forces \cite{Nottale2011}.   However, in Fig's~\ref{fractalflower}-\ref{pod}, we see the emergence of $\approx 10nm$ scale structures, in addition to an upper limit, which in practice appears closer to $\approx 20 \mu m$ under the conditions in this work.  We note that a two-scale structure is almost universal in diffusion limited growth processes such as plant cells, with cellulose nano-fibres forming the fundamental building block of a structural scaffold in the cell wall \cite{Turner2011}.

It appears that as a molecular scale fractal structure evolves, it reaches a critical point, resulting in the emergence of $nm$-scale structures.  This smaller scale of assembly may be explained by van der Waals forces playing a synergistic role alongside the macroscopic quantum potential (Eq.\ref{eq.QMG}) at the $nm$-scale.  A more speculative, additional explanation lies in a second synergistic (attractive) quantum potential originating from Casimir forces, associated with the quantum vacuum itself.  For a detailed treatment of this topic we refer the reader to theoretical studies by Simpson \cite{Simpson2015} and references therein.  Simpson's thesis suggests that Casimir forces cannot be considered in isolation in an inhomogeneous medium.  Relating the ideas developed by Simpson to our approach, the electromagnetic field associated with the quantum vacuum is fundamentally coupled to the fractal molecular scale medium, with quantization of the coupled system creating a polariton ($\psi_{pol}$).   

As in the case for charges (Eq.\ref{eq.68}), at a critical point (dictated by polariton wavelength and fractal geometry), we see transformation of quanta of polariton fractal fluctuations $\psi_{pol} = \sqrt{\rho_{pol}} \times e^{{i A_{pol}}/\hbar}$ (where $A_{pol}$ is a microscopic action), into macroscopic fluctuations, represented by a macroscopic wave function $\psi_{POL}$

\beq
\displaystyle\sum\limits_{n=1}^N\psi_{d_{pol}} \to  \psi_{POL} = \sqrt{\rho_{POL}}\times e^{i A_{POL}/2\Ds},
\label{Polpsi}
\eeq
where $A_{POL}$ is a macroscopic action and 

\beq
Q_{{POL}}  = -2{\Ds}^2 \, \frac{\Delta \sqrt{\rho_{POL}}}{\sqrt{\rho_{POL}}},
\label{eq.QMG2}
\eeq

is its associated MQP, which contributes to the emergence of structure at the $nm$ scale along with the charge induced MQP (Eq.\ref{eq.QMG}) and so is added to the existing Euler and continuity equations (Eq's \ref{eq.CMGEuler}-\ref{eq.CMGcont})

\beq
\frac{\d V_{N}}{\d t} + V_{N}. \nabla V_{N} =- \frac{\nabla \phi}{m} -\frac{\nabla Q_N}{m}-\l(\frac{\d V_{POL}}{\d t} + V_{POL}. \nabla V_{POL} + \frac{\nabla Q_{POL}}{m}\r),
\label{eq.CMGEuler2}
\eeq
\beq
\frac{\d \rho_N }{ \d t} + \text{div}(\rho_N V_N)=-\frac{\d \rho_{POL} }{ \d t} - \text{div}(\rho_{POL} V_{POL}),
\label{eq.CMGcont2}
\eeq

giving a new macroscopic Schr\"odinger equation incorporating both $\psi_N$ and $\psi_{POL}$
\beq
{\Ds}^2 \Delta \psi_N + i {\Ds} \frac{\partial\psi_N}{\partial t} - \phi\; \psi_N = -\Ds^2 \Delta \psi_{POL} - i {\Ds} \frac{ \d \psi_{POL}}{\d t}+\phi \; \psi_{{POL}}\\
\label{eq.SchrodingerC2}.
\eeq

At the 10 $nm$ scale we suggest that $Q_{POL}$ represents a small but significant force $F$, which when combined with $Q_N$, leads to the emergence of the nano-fibres.   However, since $F$ falls off rapidly with distance $d$ ($F = 1/d^4$ \cite{Simpson2015}), this constrains their size.  However, given favourable conditions (e.g., lower temperatures), it is theoretically possible that larger scales of nano-fibre may emerge. Given the declining force with increasing $d$, beyond this first scale of assembly, $Q_{POL}$ is expected to play a subordinate role in the larger 10-20 $ \mu m$ scale structures (Fig's  \ref{Plant structures}-\ref{pod}).  However, we suggest that quantum vacuum fluctuations still play a significant role (as part of collective environmental fluctuations) in bifurcation processes at the $nm$ scale leading to fractal assembly of nano-fibres into larger scale structures.

Our work suggests that the continued growth of structure beyond the $nm$-scale in Fig's  \ref{Plant structures}-\ref{pod} is driven by charge density $\rho$ and frequency $\omega$ of fluctuations $\Ds$, which determine the strength and impact of the emergent MQP (Eq.\ref{eq.QMG}) within Eq.\ref{eq.SchrodingerC}.  The associated fractal network of charge fluctuations plays a key role in the transition from mesoscale to macroscale structures.  For example, as $\rho$ increases, charge induced channels between nano-fibres offer an energetically favourable fractal network of paths for assembly, leading to more spatially coherent structures.   This is reflected in a decrease in entropy and $D_F$.  We see this in a transition from dendrites found in fractal structures (Fig~\ref{Plant structures}f) to leaf (Fig~\ref{Leaf}), or segmented flower structures (Fig~\ref{fractalflower}). With increased levels of charge, we expect closure into more symmetric structures such as that illustrated in Fig~\ref{Plant structures}c.

As charge and the associated field strength increases further, we expect these symmetric structures to close in on themselves, a process we observe first in hemispherical structures (Fig~\ref{Plant structures}d) and subsequently pod-like structures (Fig~\ref{Plant structures}e and Fig~\ref{pod}).  In theory, given sufficient charge, completely symmetric cell-like structures should emerge.  However, under Stage 1 conditions, charge was insufficient to support this.  We tested this hypothesis in Stage 2, by varying levels of $CO_2$ through nitrogen and $CO_2$ rich environments.

\begin{figure}[!ht]
\begin{center}
\includegraphics[width=16cm]{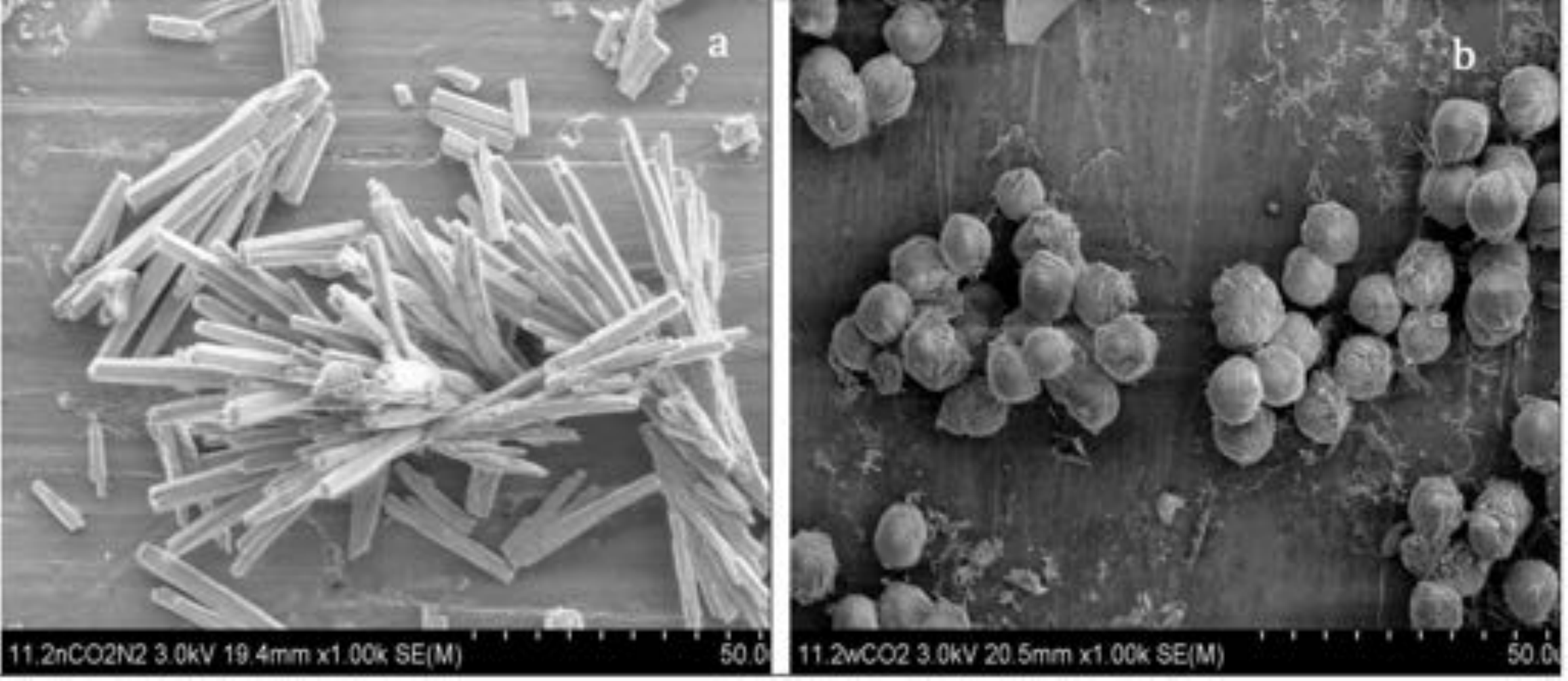} %
\caption{\small{$$ Crystals and cells formed in nitrogen (a) and $CO_2$ (b).}}
\label{cells}
\end{center}
\end{figure}

\begin{figure}[!ht]
\begin{center}
\includegraphics[width=16cm]{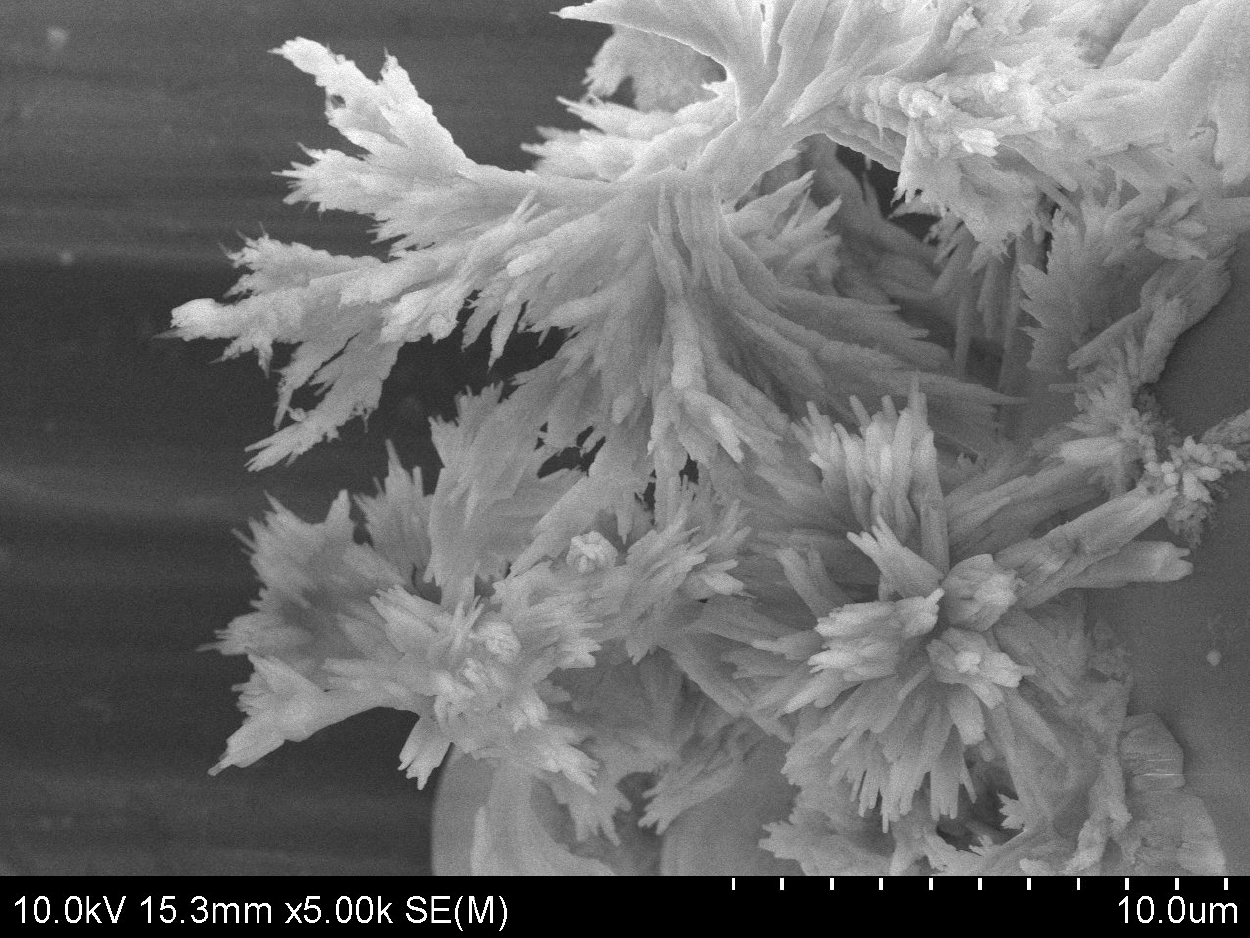} %
\caption{\small{$$ Rare fractal arrangement of crystals formed in nitrogen.}}
\label{fractalcrystals}
\end{center}
\end{figure}

\begin{figure}[!ht]
\begin{center}
\includegraphics[width=16cm]{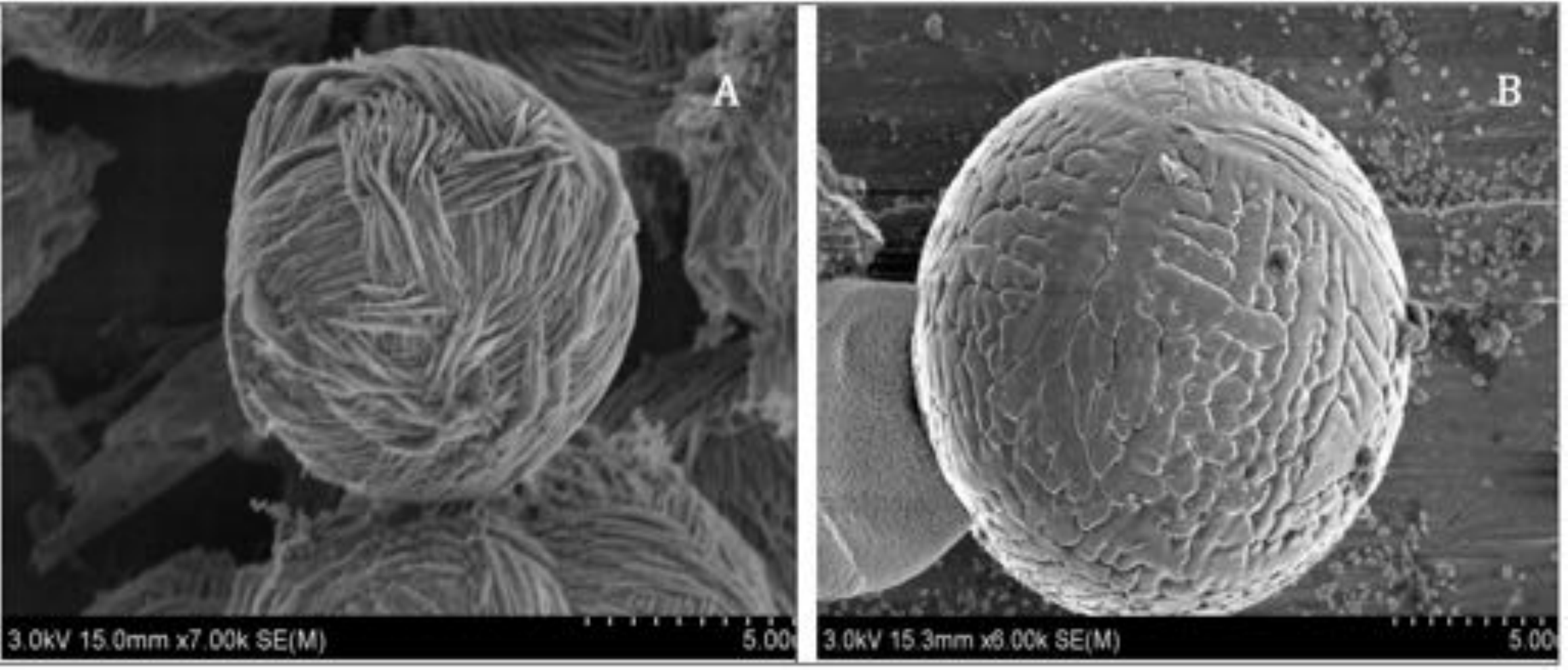} %
\caption{\small{$$ Spheres formed in elevated $CO_2$ at $\approx 18^oC$(a) and ambient $CO_2$ levels at $4^oC$ (b).}}
\label{spheres}
\end{center}
\end{figure}

The results were striking.  Elevated levels of $CO_2$ (maximum charge) led to a monoculture of spherical, cell like structures (Fig~\ref{cells}b).  By contrast, when $CO_2$ was minimised by purging the environment with nitrogen, we observed pure crystalline structures.  Fig~\ref{cells}a represents the form and scale of most of the material ($\approx 5 \mu m$ cross section x $30-50 \mu m$ length). However, in addition we observed the rare occurence of a fractal crystal (Fig~\ref{fractalcrystals}), with crystal structures right down to the $nm$ scale.  

These results indicate that as $ \rho\rightarrow 0$, molecules, unhindered by repulsive charges, are permitted to form a crystal lattice, its precise structure determined by atomic/molecular structure and external boundary conditions.  In the case of fractal crystalline structures (Fig~\ref{fractalcrystals}), we suggest that there was just enough charge to disrupt the formation of larger scale crystals, with environmental fluctuations leading to the emergence of a fractal assembly of smaller scale crystals.  These crystalline structures showed very clean surfaces, reflecting a clear difference in morphology compared to the non-crystalline structures in Fig's~\ref{fractalflower}-\ref{pod}.  Here, a fuzzy surface at higher resolutions (Fig ~\ref{pod}b), is suggestive of a fractal molecular structure, which could not be resolved in detail with the FE-SEM.  These findings appear to confirm the critical role of charge in the emergence of the types of $nm$ scale structure observed in Fig's ~\ref{fractalflower}-\ref{pod}.

Interestingly, alongside a range of structures, we also observed the emergence of completely spherical structures in the Stage 3 trial (ambient levels of $CO_2$) at $4^o$C. However, there was a significant difference in the mesoscale structure compared to results at elevated $CO_2$. Fig~\ref{spheres}a shows the detailed structure of a sphere from Fig~\ref{cells}b grown under elevated $CO_2$ conditions, revealing a significant increase in the size of nano-fibres ($\approx 250 nm$ diameter and $\approx 3 \mu m$ length) relative to structures grown at ambient $CO_2$ concentrations (Fig's \ref{fractalflower}-\ref{pod}).  This suggests that increased charge density $\rho$ leads to an increase in the $Q_N$ potential (Eq's \ref{eq.QMG2}-\ref{eq.SchrodingerC2}). The sphere grown at $4^oC$ Fig~\ref{spheres}b shows a further increase in the smallest scale fibres ($\approx 1\mu m$ diameter and up to $5 \mu m$ in length).  This increase in size combined with reduced environmental fluctuations leads to a low $D_F$ fibril structure during assembly into a spherical morphology.  Fibril diameter is now at the same scale as the pod wall thickness in Fig~\ref{pod}, suggesting the sphere may be constructed from a single layer of fibrils. Whilst this remains to be confirmed, the scale of the fibrils suggests that despite the reduction in $\rho$, lower $T$ (reduced environmental monitoring) permits the emergence of larger molecular-scale fractal networks, with Casimir forces linked to $Q_{POL}$ playing a more significant role within the system compared to Fig~\ref{spheres}a.

We do not yet have conclusive evidence to support the hypothesis of a fractal architecture at the molecular scale. However, the differences in fuzzy surface topology of $nm$-fibrils in Fig's \ref{fractalflower}-\ref{pod} and the clean surfaced crystalline structures in Fig~\ref {fractalcrystals} is strongly indicative of a fractal molecular structure in the former.  In addition, the hypothesis that fractal order exists at the molecular scale and can lead to quantum coherence at the $nm$ scale is supported by Quochi {\it{et al}} \cite{Quochi2005,Quochi2010}, where CRL was reported in organic nano-fibres.  At another level, we note that CRL has been reported in inhomogeneous nano-fibre suspensions \cite{Lee2015}, this contrasts with ordered systems in the same work, where CRL disappeared, suggesting that a continuous structure from molecular to $nm$ scales is not essential for macroscopic coherence.  

To confirm fractal order at molecular to $\mu m$ scales, future work is planned to determine charge distribution following an approach reported by Fratini {\it{et al}} \cite{Fratini2010} using scanning synchrotron radiation X-ray micro diffraction and a charge coupled area detector.  To complement this work, studies similar to those on cellulose in plants using XRD and NMR \cite{Testova2014} will be used to confirm the absence (or presence) of crystalline structure in $nm$ scale structures.  

\textbf{Additional observations}

In exploring the opportunity to improve the resolution of FE-SEM images of structures, we varied the electron beam voltage.  The default setting was 3kv.  As voltage increased significantly, we made an unexpected observation of high levels of fluorescence in some of the plant-like structures we have reported here.  The fluorescence looked remarkably like that observed in CRL.  

\begin{figure}[!ht]
\begin{center}
\includegraphics[width=10cm]{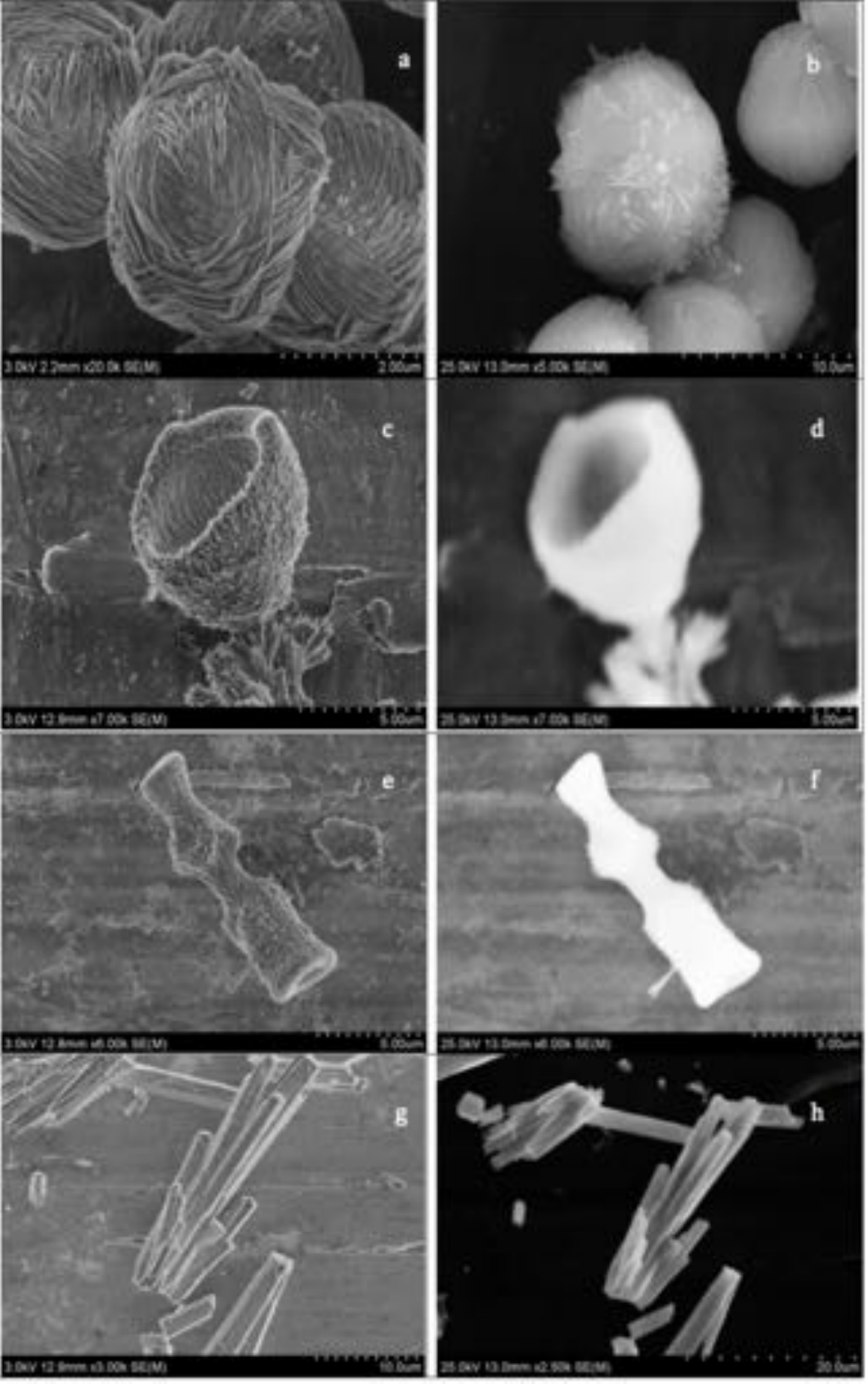} %
\caption{\small{$$ Observation of fluorescence in non-crystalline materials.}}
\label{flores}
\end{center}
\end{figure}

In the absence of an alternative explanation for the fluorescence, we speculate (rather tentatively) that the phenomenon might be analogous with CRL, with electron fluorescence being an indicator of macroscopic quantum coherence.   As a first step in testing this idea we assessed different structures at the standard 3kv setting and at a higher level of 25kv.  This represents the equivalent of a significant increase in gain required to induce CRL: below a critical level of gain CRL is not observed \cite{Turner2015}.  In Fig~\ref{flores} we see examples from the assessment.  On the left hand side of Fig~\ref{flores} we see images of a sphere (A), chalice\footnote{resembling the chalice sponge ({\it{Heterochone calyx}})}(C), bone-like structure (E) and crystals observed with a 3kv electron beam.  On the right, the same structures are observed at 25kv.   

In the first pair of images (Fig~\ref{flores}a \& b), a distinctive fluorescence is observed in the right hand sphere at 25kv, with very little of the $nm$-scale being observable, due to this fluorescence.  In subsequent images (Fig~\ref{flores}c-f), higher levels of fluorescence are observed at 25kv in Fig~\ref{flores}d \& f, with the complete disappearance of surface detail.  

We speculate that the lower level of fluorescence in Fig~\ref{flores}b is due to better electron confinement, with the more open surface structures in Fig~\ref{flores}c-f facilitating electron escape from the structure, supporting the analogy with CRL.

In the last pair of images of ordered crystalline structure (Fig~\ref{flores}g-h), the distinctive fluorescence observed in structures with fractal mesoscopic structure is absent.  This is predicted if the fractal architecture in Fig~\ref{flores}a-f is supporting macroscopic coherence of electrons.

In considering an appropriate interpretation of these observations we need to ask if room temperature macroscopic electron coherence (aided by high vacuum in the SEM) is theoretically possible in these structures.  If we consider previous theoretical work \cite{Turner2015}, then it may indeed be the case.  Taking a first principles approach, these structures meet a key criteria for high temperature superconducting material, namely a three dimensional fractal structure, and fractal charge density distribution, which in the current experimental conditions emerges naturally to form coherent structures such as spheres.  

Previous experimental work \cite{Turner2015} has shown that $e$-pairs in the pseudo gap phase can exhibit coherence at critical temperatures $T_c$ well above room temperature.  However, in this earlier work $e$-pairs were localized and so did not contribute to conduction.  We postulated that if structures were designed optimally in a 3D fractal architecture, then high strength macroscopic quantum potentials could theoretically support delocalization and conduction at room temperatures \cite{Turner2015}.  The current work suggests that at least some of the coherent plant-like structures (particularly spheres) could meet these criteria.  This idea is supported by work on photosynthetic systems which provide conclusive evidence that relatively long-lived quantum coherent states exist at room temperature in protein complexes \cite{Engel2007,Collini2010}.

We stress that whilst these preliminary observations appear interesting, they were unexpected.  As a first step, future work will focus on confirming or refuting the possibility that the observed fluorescence in these objects reflect macroscopic coherence of electrons by measuring their $T_c$.  

\subsection*{4.2.3. Modelling macroscopic structures with a macroscopic Schr\"odinger equation}
 
\begin{figure}[!ht]
\begin{center}
\includegraphics[width=10cm]{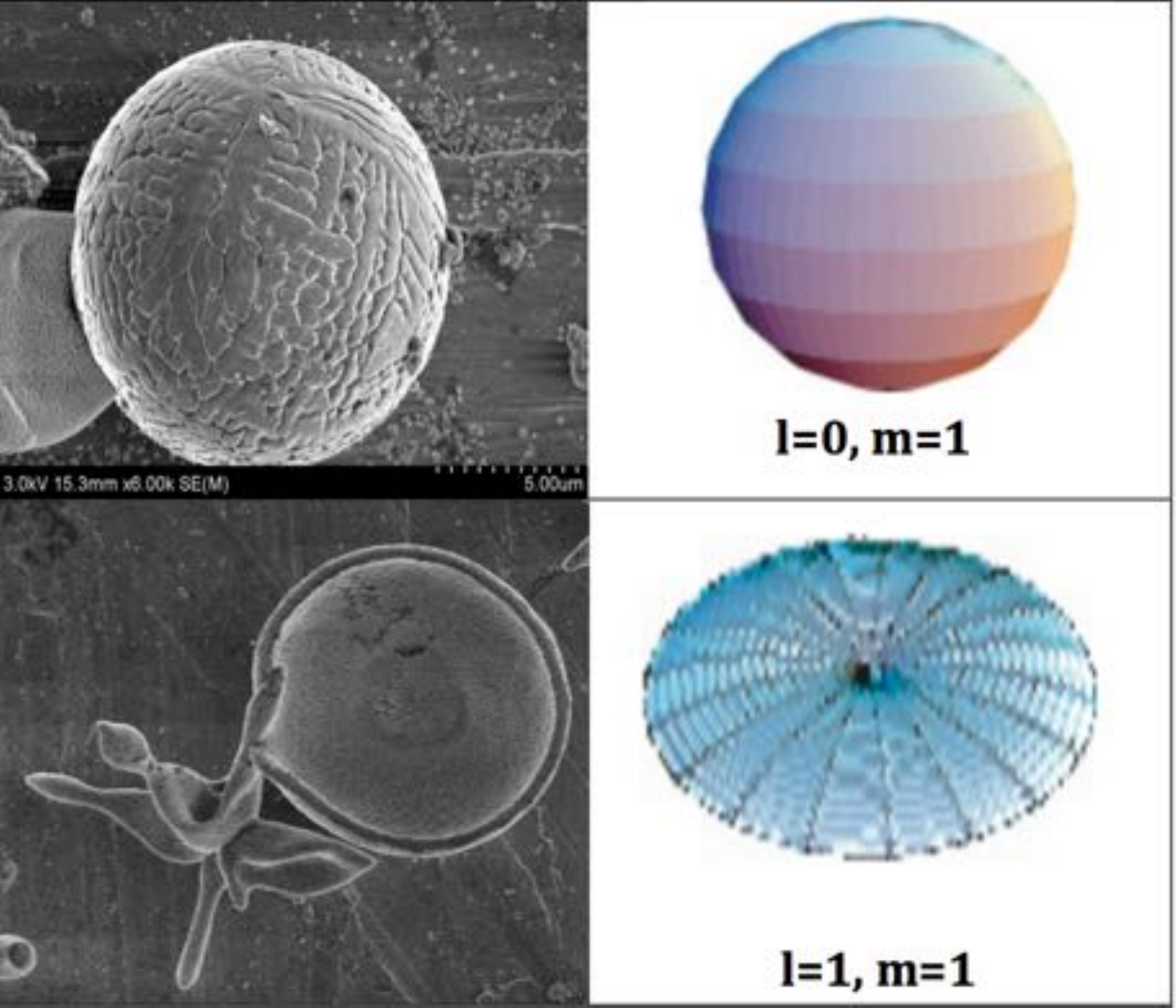} %
\caption{\small{ Examples of structures observed compared to quantized morphologies for ejection processes associated with planetary nebulae determined by quantum numbers l and m.}}
\label{Models1}
\end{center}
\end{figure}

\begin{figure}[!ht]
\begin{center}
\includegraphics[width=10cm]{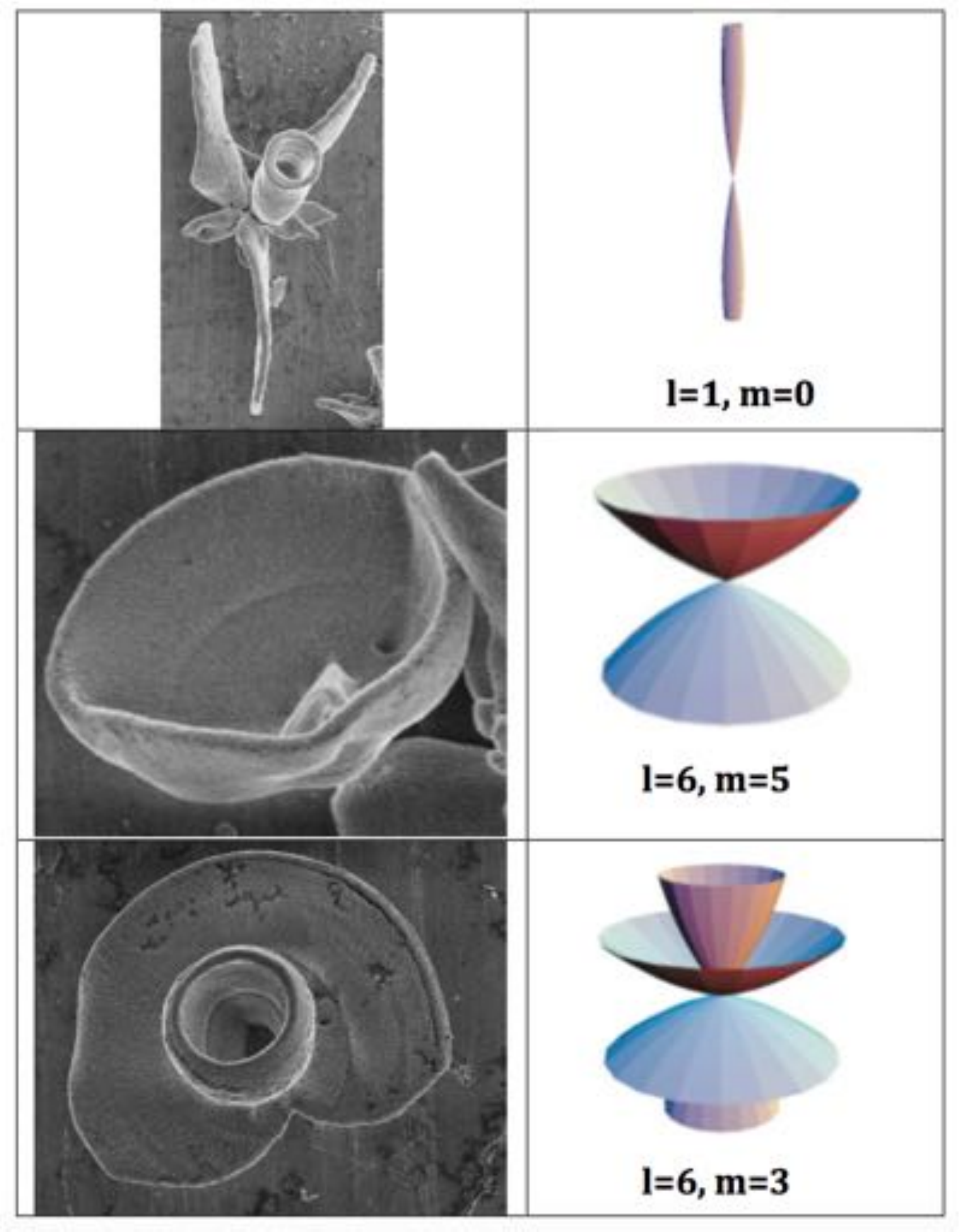} %
\caption{\small{ Examples of structures observed compared to quantized morphologies for ejection processes associated with planetary nebulae determined by quantum numbers l and m.}}
\label{Models2}
\end{center}
\end{figure}

Having considered the detail, which leads to the emergence of a macroscopic quantum system, we suggest that a number of the structures observed share common features and processes with planetary nebulae (stars that eject their outer shells) reported by da Rocha and Nottale \cite{da Rocha2003a,da Rocha2003b}.  In this earlier work, the chaotic motion of ejected material was modelled using a macroscopic Schr\"odinger equation, describing growth from a centre, corresponding to an outgoing spherical probability wave, having well defined angular solutions $\psi (\theta, \phi)$.  Their squared modulus $P=|\psi^2|$ is identified with a probability distribution of angles characterized by the existence of maxima and mimima.  These in turn are dependent on the quantized values of the square of angular momentum $L^2$, determined by the quantum number $l$ and its projection $L_z$ on axis $z$, which is characterized by the quantum number $m$.  This means that $L^2$ and $L_z$ can only take specific values proportional to these quantum numbers, which are integers, allowing the prediction of discrete morphologies.  We see striking parallels between examples of the two sets of structures illustrated in Fig's~\ref{Models1} and \ref{Models2}, but with the caveat that the inorganic structures are constrained by growth on a plate, whilst planetary nebulae in Fig~\ref{Models2} show a double ejection process in space.

\begin{figure}[!ht]
\begin{center}
\includegraphics[width=12cm]{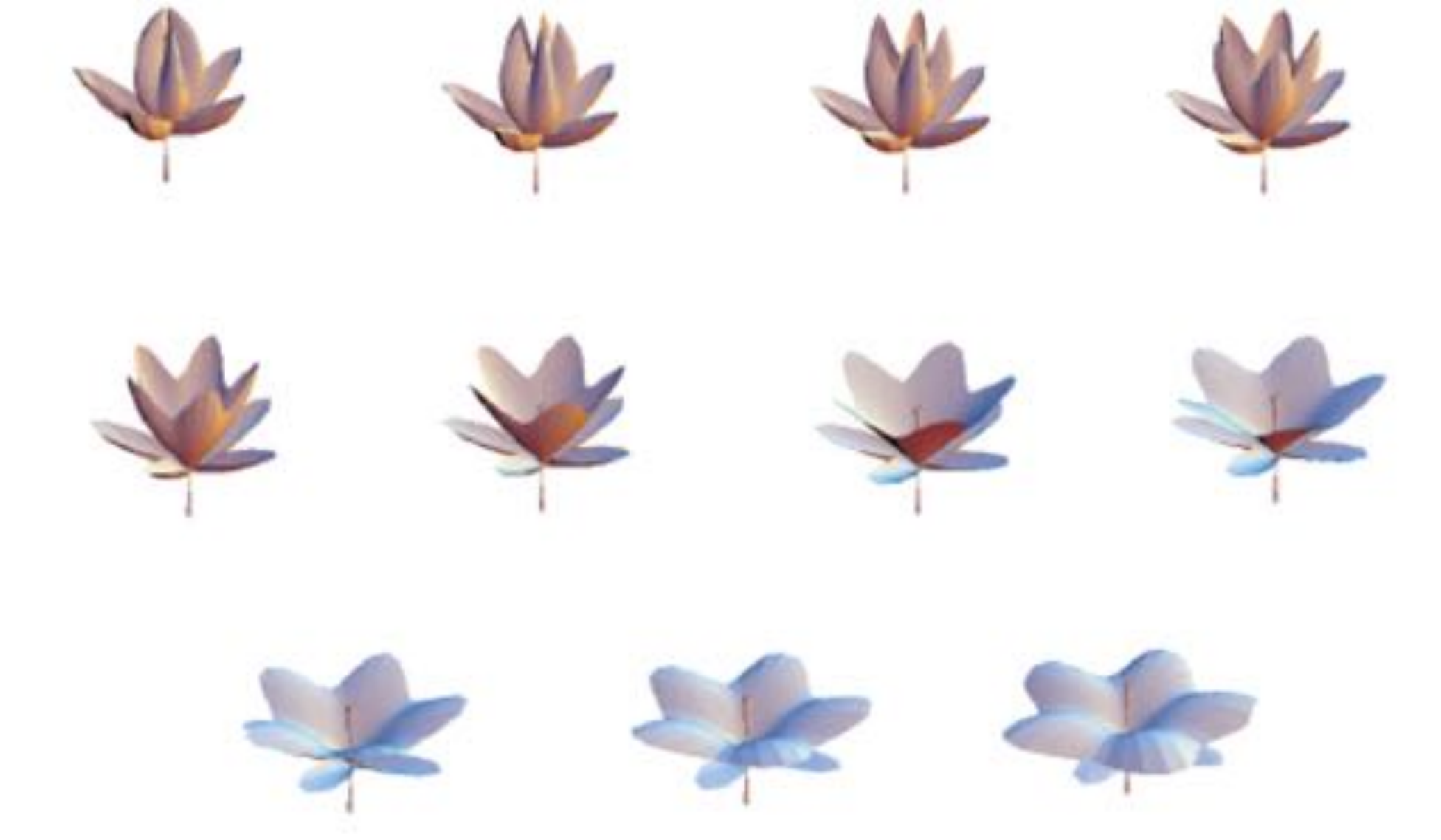} %
\caption{\small{Morphogenesis of a flower-like structure, solution of a time-dependent Schr\"odinger equation describing a growth process from a centre ($l=5, m=0$). The `petals', `sepals' and `stamen' are traced along angles of maximal probability density. We note that these different structures are not built from different equations.  The whole structure originates from a single equation.  By varying the force of `tension' we see a sequence of images simulating the opening of the flower, reproduced from Nottale and Auffray \cite{Nottale2008}.}}
\label{Morphogenesis}
\end{center}
\end{figure}

In previous comparisons between this process and plant morphogenesis \cite{Nottale2008}, modelling of the growth of flower-like structures, with morphologies evolving along angles of maximal probability has been described. In the case of flower-like structures, spherical symmetry is broken and one jumps to discrete cylindrical symmetry.  In the simplest case, a periodic quantization of angle $\theta$ (measured by an additional quantum number $k$), gives rise to a segmented structure (discretized `petals').  In addition, there is a discrete symmetry breaking along axis $z$ linked to orientation (separation of `up' and `down' due to gravity, growth from a stem). This results in successive structures illustrated in Fig~\ref{Morphogenesis}, indicating the evolution of a range of possible outcomes, which offers insight into the mechanism driving a segmented flower-like structure such as that observed in Fig~\ref{fractalflower}.  

We note that Fig~\ref{Morphogenesis} gives an example of just one possible scenario. Depending on the potential, on the boundary conditions and on the symmetry conditions, a large family of solutions (structures) can be obtained when conditions for the quantum-type regime are fulfilled.

This work offers convincing evidence to support the hypothesis that the emergence of inorganic, plant-like structures reported here and by Norrduin {\it{et al}} \cite{Norrduin2013} can be explained within the context of a macroscopic quantum-type process,  induced by a fractal network of charges.  

The process is  characterized by competing systems, which can be controlled by either reducing external fluctuations $H_E$ (by decreasing $T$) or increasing internal forces $H_S$ (by increasing $\rho$).  However, we note that with the exception of extreme conditions illustrated in Stage 2 (Fig~\ref{cells}), control is not precise, but impacts on the probability of certain outcomes.  This principle is reflected in the probability of the emergence of fractal structures (reflecting macroscopic pointer states created by the fractal velocity field), which increased with distance from the solution on the aluminium plates, as the result of a proton gradient.  In this instance,  decreasing $\rho \rightarrow$ increasing probability of less ordered (fractal) morphologies.

The principle is further illustrated in ice formation, in a process analogous with our laboratory work and that by Norrduin {\it{et al}} \cite{Norrduin2013}.  Ice structures range from needle crystals to fractal snowflakes and beyond to various ice flower morphologies and spheres (hailstones).  We suggest that structure is driven in a similar mechanism by levels of ionization and temperature, with the most extreme conditions (e.g., plasma induced ionization during a thunderstorm) leading to hailstones.  This theory could be easily tested by determining the influence of $T$ and $\rho$ on the emergence of different ice morphologies.

Such a mechanism also has its equivalence in ionized gases, which at sufficiently high levels of charge density and localized zones of charge differential, leads to fractal patterns of discharge (lightning).  However, as $\rho$ increases, we see a transition from fractal to sheet to ball morphologies (`ball lightning') in a process analogous with that observed in Fig~\ref{cells}B.  

\subsection*{4.2.4. Plant-scale systems}

The work by Norrduin {\it{et al}} \cite{Norrduin2013}, and that reported here parallels in a very striking way, the emergence of a range of structures found in the plant kingdom, but now on the scale of cells, (typically $10 \mu m - 25\mu m$).  As stated previously, this scale is explained by a natural symmetry breaking as gravitational forces exceed van der Waals forces.  Biological systems address this constraint on scale, with growth processes evolving via a replicative cellular structure to generate larger scale structures. 

In order to translate theory and experimental results on inorganic plant-like structures to their biological equivalent, we now consider the emergence of both mesoscale structures (in plant cells) and more complex, multicellular structures.  At the heart of this process, we need to explain how the wide variety of structures we observed emerging spontaneously under standard atmospheric concentrations of $CO_2$ in inorganic plant-like systems is more precisely (and repeatably) controlled in real plants.  

\subsection*{4.2.4.1. Mesoscale structures}

As a first step we consider mesoscale cellulose structures that form the structural scaffold of plant cell walls \cite{Turner2011}.  In most higher level plants, cellulose chains are extruded from a 6 x 6 arrangement of rosettes (the cellulose synthase complex) creating short, 36 chain crystalline units of cross section $\approx$ 3-6 $nm$ \cite{Peciulyte2015}.  The subsequent assembly of these crystalline units into nano-fibrils and their patterning in the cell wall is dictated by microtubule bundles \cite{Derbyshire2015}, the details of which we consider later in this section.  

To date the reason for relatively short crystalline units has not been satisfactorily explained.  However, based on the principles established in this work, it seems likely that they result from charge-induced disruption of the crystal lattice during the extrusion process from the cellulose synthase complex.  The result is predicted to be a fractal assembly of crystalline units (induced by environmental fluctuations), interspersed with amorphous cellulose and hemicellulose chains, which aggregate to form $\approx$10-20 $nm$ diameter composite nano-fibres \cite{Turner2011,Peciulyte2015}.  However, it seems logical that charge density will determine the precise internal composition of these nano-fibres (the source of considerable debate), along with their subsequent assembly into larger micron scale structures in the cell.  This means that in some circumstances, it is possible that at low levels of charge, larger crystalline structures, up to the scale of the nanofibril, may exist within some species of plant (or at specific positions within the plant).  This level of detail may not be resolved in averaging techniques associated with XRD or NMR analysis \cite{Peciulyte2015}. The hypothesis is supported by observations in aquatic plants such as Valonia (a genus of green algae in the Valoniaceae family), where larger scale, pure crystalline cellulose structures ($\approx$ 50 $nm$ cross section and > 500 $nm$ in length) have been observed \cite{Sugiyama1984}.  In this instance, if we accept that $CO_2$ may have some influence on charge density, then lower levels of $CO_2$ in aquatic environments (compared to land based plants which are able to absorb higher levels of atmospheric $CO_2$), could at least partially explain lower levels of charge induced disruption of the crystal lattice.

\begin{figure}[!ht]
\begin{center}
\includegraphics[width=16cm]{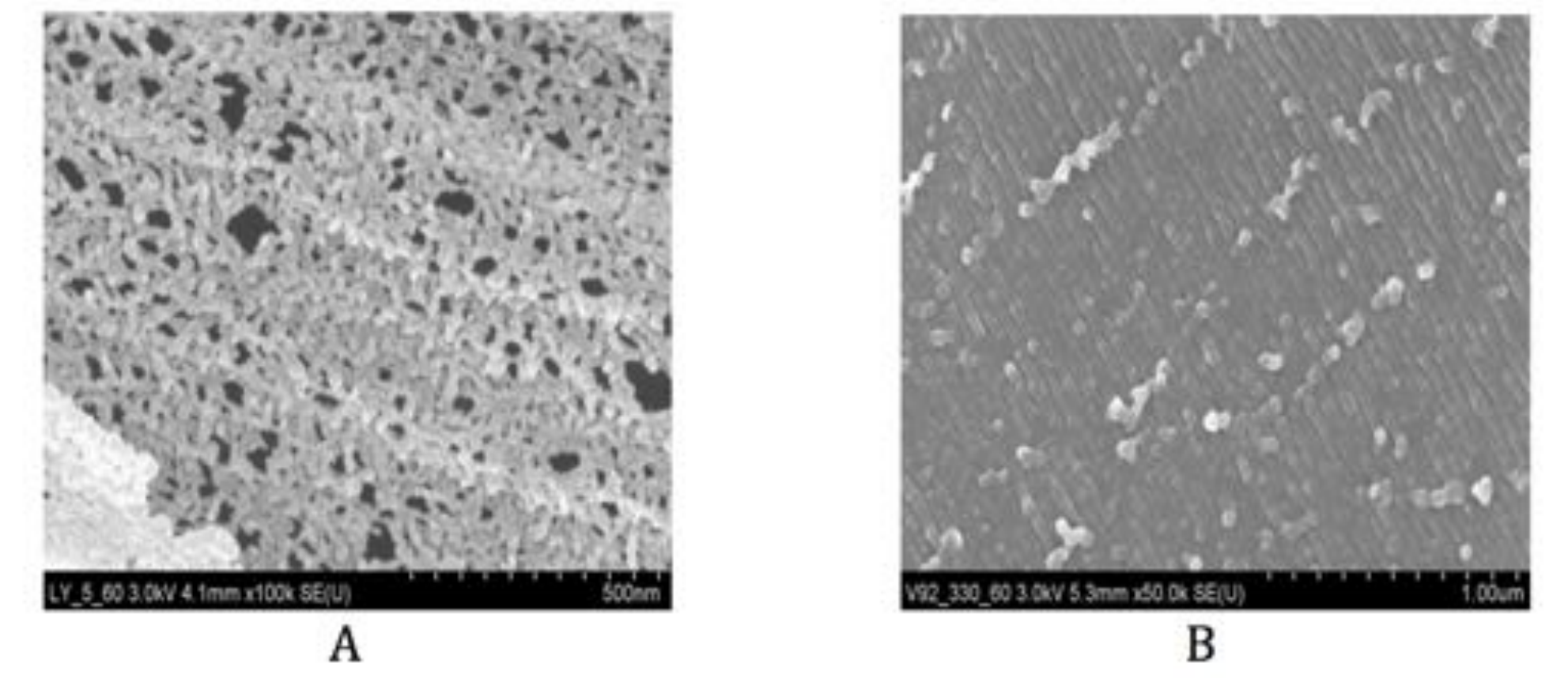} %
\caption{\small{Mesoscale structure of crystalline cellulose in the S2 layer (A) and S1 layer (B) of a {\it{Eucalyptus grandis}} fibre \cite{Turner2011}.}}
\label{Nanocrystals}
\end{center}
\end{figure}

These principles are also valid at higher scales of assembly.  For example, the internal mesoscale fractal architecture of the structures observed in Fig's~\ref{fractalflower}-\ref{pod} has its analog in the fractal assembly of cellulose nano-fibres observed in the S2 layer of a {\it{Eucalyptus grandis}} cell wall illustrated in Fig~\ref{Nanocrystals}A.  However, within the same cell  \cite{Turner2011}, we see alternative structures, such as the nematic assembly of cellulose nanofibres (Fig~\ref{Nanocrystals}B) found in the S1 layer, which we would expect under conditions where charge density drops significantly.  Clearly, in these circumstances, atmospheric $CO_2$ concentration, which remains constant, cannot be the sole factor dictating these structures.  We therefore require a mechanism to explain how the plant genome has evolved to tightly control levels of ionisation, which can be changed in an instant, resulting in the emergence of the diverse arrangements of nano-fibres within a cell needed to meet the structural requirements of the plant.

\begin{figure}[!ht]
\begin{center}
\includegraphics[width=16cm]{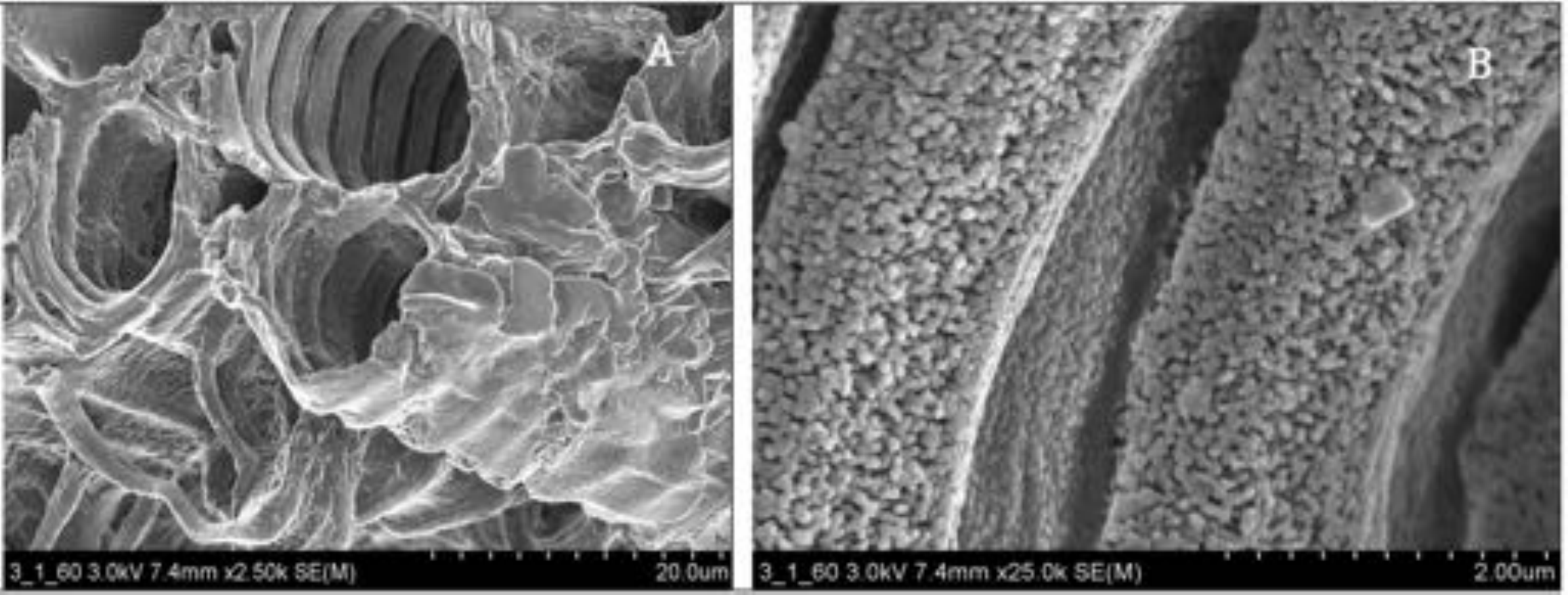} %
\caption{\small{Tracheary elements in {\it{Aradidopsis thaliana}} (A) alongisde a higher magnification illustrating the detailed internal structure of an individual cell's rib-like circumferential secondary wall thickening (B)}}
\label{Arabidopsis}
\end{center}
\end{figure}

To give an idea of how structures in plants are more precisely controlled we consider recent work on the impact of charged macromolecules (proteins) on the emergence of different structures associated with the deposition of cellulose in secondary wall thickening in tracheary elements of {\it{Arabidopsis thaliana}} \cite{Derbyshire2015}.  A typical example of the cell's rib-like annular secondary wall thickening is illustrated in Fig~\ref{Arabidopsis}.  The $\approx$1.5 $\mu m$ cross section discrete structures appear to be composed of a fractal arrangement of cellulose nanostructures, which contrast with the more uniform distribution of secondary wall structure of cellulose nano fibres observed in {\it{Eucalyptus grandis}} in Fig~\ref{Nanocrystals}.

As discussed previously with reference to the deposition of cellulose nanofibres, the organization of secondary wall thickening in tracheary elements (TE's) varies according to the patterning of microtubule bundles that guide wall deposition.  Derbyshire et al \cite{Derbyshire2015} has shown that whilst microtubules provide the guide for deposition of secondary cell wall thickening, the overall patterning of these guiding microtubules is regulated by specific microtubule-associated proteins (MAPs) and protein complexes. 

The study on individual stem cells in vitro \cite{Derbyshire2015} identified 605 different proteins that associate with microtubules in the narrow window of time in which these cell wall thickenings are deposited in the secondary cell wall. Through control of gene expression, different sets of MAP's clearly influenced patterning of secondary cell wall deposition.  Three different types of structures (a continous spiral, a more inter-connected reticulated pattern and finally a more dense packing interspersed with a pitted structure), were directly correlated with specific proteins (see Fig 4a-d \cite{Derbyshire2015}).  From this work it becomes apparent that the overall charge density, associated with specific combinations of proteins (protein complexes), controls the assembly of cellulose into specific structures with a level of precision and repeatability that is impossible to achieve in the inorganic plant like structures using $CO_2$ concentration described in experimental work in the current paper.

We conclude that during the transcription of DNA through RNA to a vast array of different proteins (which are a direct reflection of the genetic code), the subsequent assembly of protein complexes is tightly controlled to effect subtle changes in charge density (and therefore the `average' charge) on the protein complex, which in effect generate their own macroscopic quantum potentials\footnote{This concept is independently supported by recent  theoretical work \cite{Vattay2015}, suggesting that complex protein structures have evolved as a key component of biological systems, precisely because their complex structures exhibit macroscopic quantum properties, which play a key role in biochemical electronic processes.}.  These protein complexes, which attach to the microtubule, control not only the patterning of the microtubule but also offer a very precise mechanism for dosing the charge at the point of cellulose assembly, thus dictating its final structure (gene expression) at the point of interaction with the cellulose synthase complex.  However, the transcription of genes into gene products (protein and protein complexes) represents just one aspect of the control of charge density. To get the full picture we also need to consider the impact of a range of different ions within the plant which can potentially interact in the gene transcription and post-translational modification of proteins to influence molecular assembly.  An obvious example that requires more work within the context of the laboratory studies discussed in the present paper includes the potential for generic protonation linked to $CO_2$ concentration.  This can be expected to reveal itself in a classic genotype/environment interaction, an example of which we consider at the plant scale in Section 4.2.4.3.

A further example to illustrate the importance of charge includes the phosphoryl group $PO_3^{2-}$. Phosphorylation alters the charge on a number of protein complexes, which leads to conformational change, such as a fold in the structure, resulting in a change in their function and activity. Many of these changes can have an indirect but important impact on cell formation.  As an example, phosphorylation of $Na^+$/$K^+$-ATPase influences the transport of sodium ($Na^+$) and potassium ($K^+$) ions across the cell membrane in osmoregulation, which in itself has a profound impact on the cell's response during the growth (assembly) process.

\subsection*{4.2.4.2. Cell duplication}

In considering cell duplication, we find a precedent for its description in macroscopic quantum processes.   In previous work \cite{Nottale2008,Nottale2011,da Rocha2003a,Nottale1996}, an example of duplication is given of the formation of gravitational structures from a background medium of uniform mass density $\rho$. This problem has no classical solution, since no structure can form and grow in the absence of large initial fluctuations.  By contrast, in the present quantum-like approach, the stationary Schr\"odinger equation for an harmonic oscillator potential (which is the gravitational potential created locally by a medium of constant density) does have confined stationary solutions.  Solving for the Poisson equation yields a harmonic oscillator gravitational potential $\varphi(r)=2\pi G \rho r^2/3$, and the motion equation becomes the Schr\"odinger equation for a particle in a 3D isotropic harmonic oscillator potential

\beq
\Ds \Delta \psi + i \Ds \frac{\partial \psi}{\partial t}- \frac{\pi}{3} G \rho r^2 \psi =0,
\label{harmonicO}
\eeq

with frequency
 
 \beq
 \omega =2 \sqrt \frac{\pi G \rho}{3}.
 \eeq
 
The stationary solutions \cite{Nottale1996,Landau1967} are expressed in terms of the Hermite polynomials ${\cal H}_{n}$,
 
 \beq
 R(x,y,z)\, \alpha\, exp \l( -\frac{1}{2} \frac{r^2}{r^2_0}\r) { {\cal H}_{n}}_{x} \l( \frac{x}{r_0}\r)  { {\cal H}_{n}}_{y}  \l( \frac{y}{r_0}\r)  { {\cal H}_{n}}_{z}  \l( \frac{z}{r_0}\r),
 \eeq
 
 which depend on the characteristic scale
 
 \beq
 r_0 = \sqrt \frac{2\Ds}{\omega} = \Ds^{1/2} (\pi G \rho/3)^{-{1/4}}.
 \eeq

The energy over mass ratio is also quantized as 

\beq
\frac{E_n}{m}=\, 4 \Ds \sqrt \frac{\pi G \rho}{3} \l(n+ \frac{3}{2}\r).
\eeq

The main quantum number $n$ is an addition of the three independent axial quantum numbers $n=n_x +n_y +n_z$. 

\begin{figure}[!ht]
\begin{center}
\includegraphics[width=10cm]{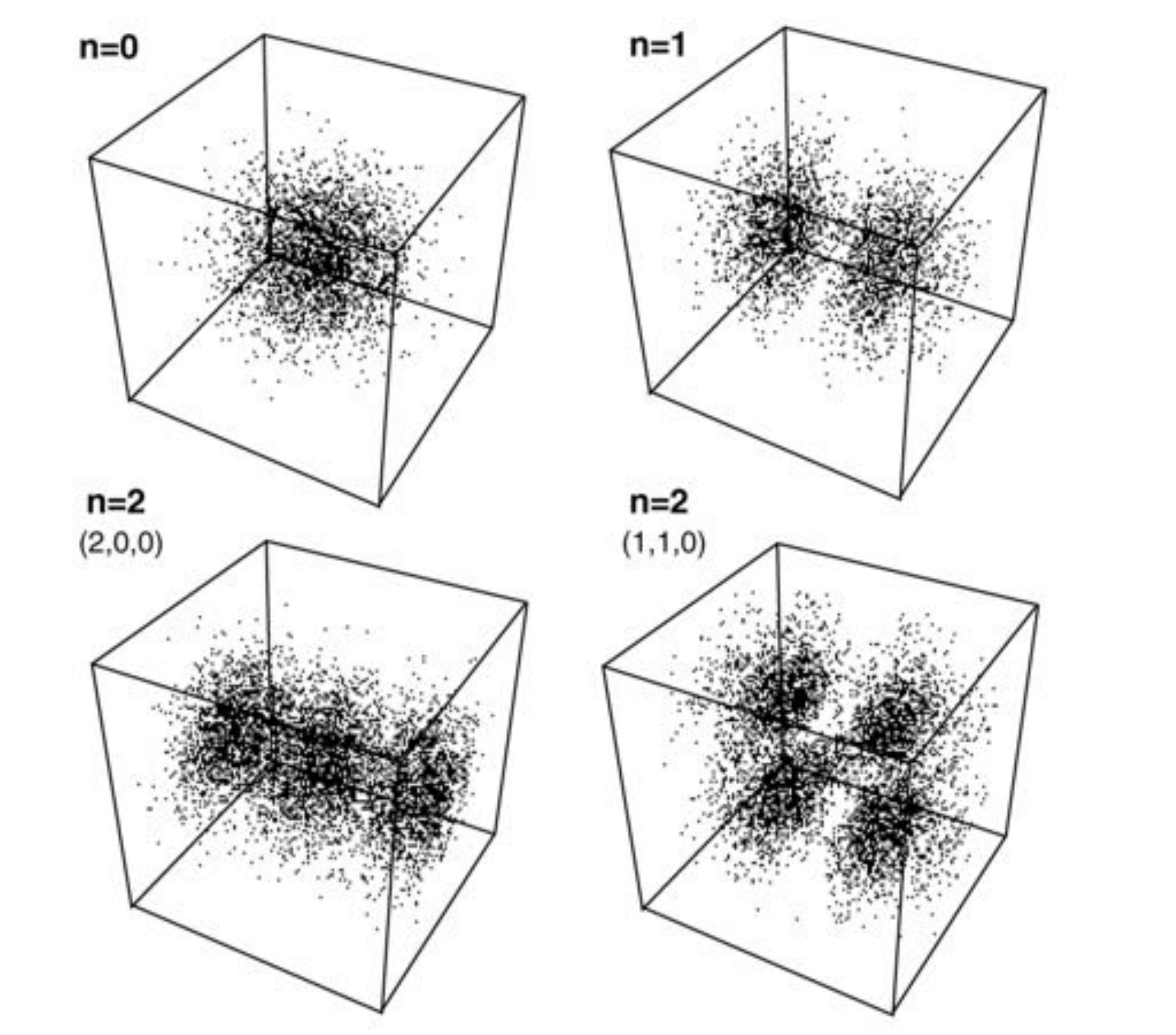} %
\caption{\small{Model of duplication, reproduced from Nottale \cite{Nottale2011}.  These solutions have been simulated by distributing points according to the probability density.  $n =1$ corresponds to the formation of binary objects (binary stars, double galaxies, binary clusters of galaxies). }}
\label{Duplication1}
\end{center}
\end{figure}

Fig~\ref{Duplication1} illustrates stationary solutions of Eq.\ref{harmonicO}.  The fundamental level or vacuum solution (the vacuum is the state of minimal energy), defined by the quantum number $n=0$, results in a one-body structure with Gaussian distribution. Subsequent solutions imply that in case of energy increase, the system will not increase its size, but will jump from a single to a double structure $n=1$, with no stable intermediate step between the two.  As energy levels increase further, the mode $n = 2$ decays into two sub-modes reflecting a chain and then a trapeze structure.  Whatever the scales in the Universe (stars, clusters of stars, galaxies, clusters of galaxies), the zones of formation show in a systematic way these kinds of structures \cite{Nottale2011}.

\begin{figure}[!ht]
\begin{center}
\includegraphics[width=12cm]{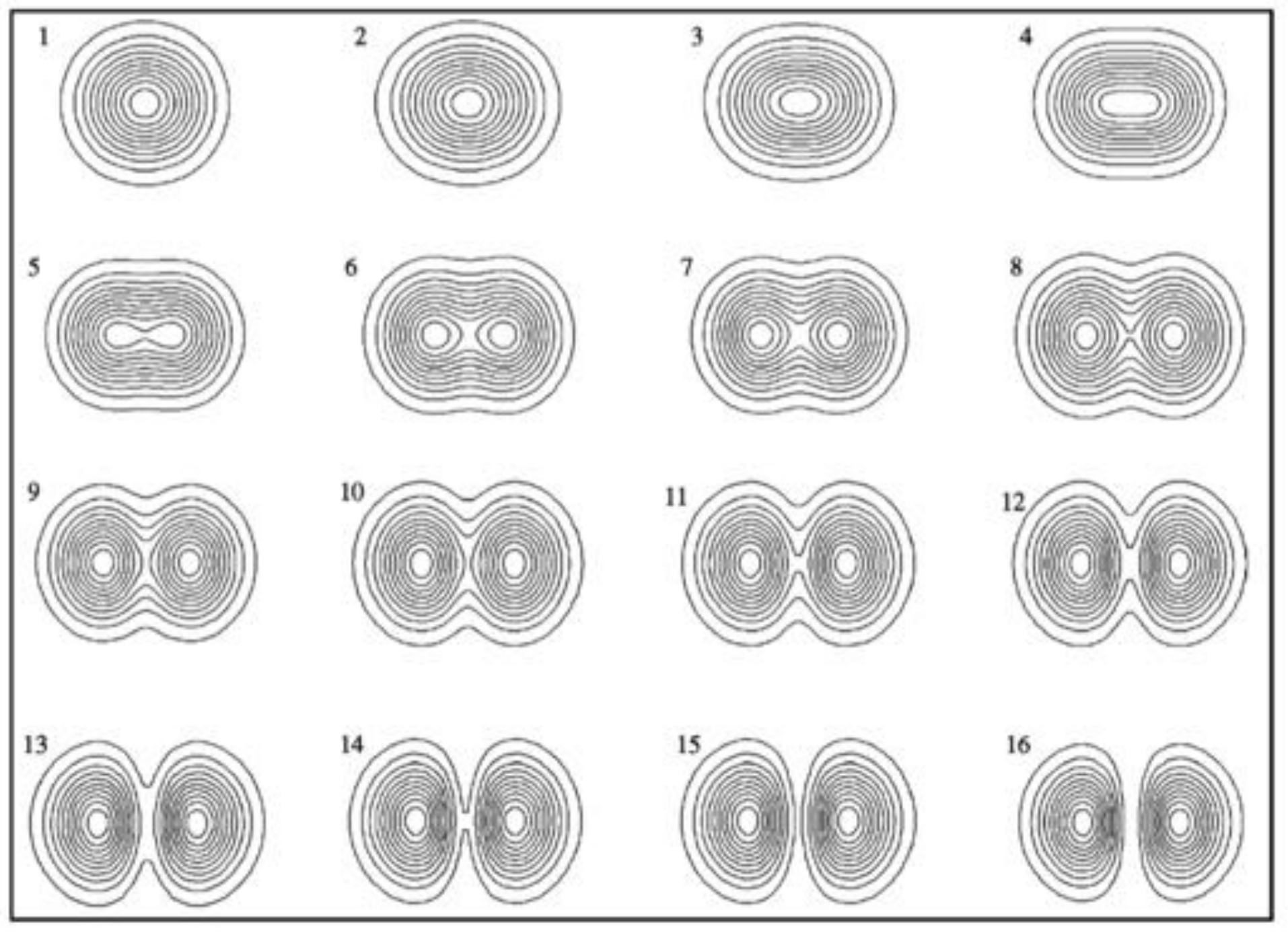} %
\caption{\small{Scale relativity model of cell duplication, reproduced from Auffray and Nottale \cite{Nottale2008}.  The successive figures give the isovalues of the density of probability for 16 time steps.  The first and last steps ($1, n=0$ and $16, n=1$) are solutions of the stationary (time-independent) Schr\"odinger equation, whilst intermediate steps are exact solutions of the time-dependent Schr\"odinger equation and therefore reflect only transient structures.  The process is similar to bifurcation (Fig~\ref{Bifurcation.pdf}), where the previous structures remain and add to themselves along a z-axis instead of disappearing as in cell duplication. }}
\label{Duplication}
\end{center}
\end{figure}

Taking an alternative perspective on successive solutions of the same time-dependent 
Schr\"odinger equation in a 3D harmonic oscillator potential, Fig~\ref{Duplication} shows how the system jumps from a one-body to a two-body structure when the jump in energy takes the quantized value $2\Ds \omega$.  Whilst this approach is particularly well suited to describe cell duplication, it is of course far from describing the complexity of true cellular division.  However, it serves as generic model for a spontaneous duplication process of quantized structures, linked to energy jumps in the presence of environmental fluctuations.

\subsection*{4.2.4.3. Multi-cell structures.}

Extending the conditions of replication outlined in Fig's~\ref{Duplication1}-~\ref{Duplication}, one creates a `tissue' of individual cells, which can be inserted in a growth equation, which once again takes a Schr\"odinger form.  Its solution yields a new, larger scale of organization.

In biological systems such as plants, growth is characterized by fractal architectures, created through bifurcation processes (Fig~\ref{Bifurcation.pdf}), which replicate structures found at the molecular to $nm$-scale, but now with assembly at the scale of cells, the cell playing the role of the `quanta of life'.  The resulting multi-scale fractal architecture creates once again, the conditions required for the formation of a complex velocity field $\widehat V$, with varying energy levels (dependent on local fractal geometries) across the entire plant.  These conditions lead to the emergence of quantized morphologies, with a range of boundary conditions dictating the wide range of coherent and fractal structures we see in living systems.  

One of the main interests of the macroscopic quantum-type approach is its capacity to make predictions about the size of the structures, which are formed from its self-organizing properties.  In some cases, this depends only on the boundary conditions, i.e., on the environment (in a biological context). Consider for example the free geodesics in a limited region of space. The classical fluid equation would yield a constant probability density (i.e., no structure), while the scale relativity description yields an equation similar to the Schr\"odinger equation for a particle in a box, which is solved in one dimension in terms of a probability density

\beq
P=|\psi|^2= \frac {2}{a} sin^2 \l(\frac {\pi \,n\,x}{a}\r) .
\eeq

The multidimensional case is a product of similar expressions for the other coordinates. One therefore obtains, at the fundamental level ($n=1$), a peaked structure whose typical size is given by its dispersion

\beq
\sigma_x= \frac {a}{2 \pi} \sqrt { \frac{\pi^2}{3} -2} \approx 0.1807a
\eeq

and therefore depends only on the size of the box \cite{Nottale2008}.

In other cases, the size depends on the fluctuation parameter $\Ds$. For example, in the harmonic oscillator solutions considered in Fig's~\ref{Bifurcation.pdf} and \ref{Duplication1}-\ref{Duplication}, the dispersion of the Maxwellian probability density distribution of the fundamental level is given by $\sigma ^2=\hbar/m\omega$, where $\omega$ is the proper frequency of oscillations, i.e., in the generalized case,

\beq
\sigma = \sqrt {\frac {\Ds}{\omega}}
\eeq

The Planck constant $\hbar$ in standard quantum physics is determined by experimental observation, then used to predict outcomes in new experiments.   We contemplate the same approach in macroscopic quantum physics, even though the value of $\Ds$, which is defined as the amplitude of mean fractal fluctuations described by Eq.\ref{diffusionrelation} (which is defined as a diffusion coefficient) is no longer universal.  

For a given system, one expects the appearance of many different effects from such a macroscopic quantum-like theory, including, interferences, quantization of energy, momentum, angular momentum, shapes, sizes, angles, etc., so that the constant $\Ds$ can be measured from any of these effects (e.g., the energy of the linear oscillator is $E_n=(2n+1) \Ds \omega)$ and then taken back to predict the size ($\sqrt {\Ds / \omega} $ for the linear oscillator) and other properties of the system under consideration. In such a case, scales will be determined by the new definition of the de Broglie length 

\beq
\lambda_{deB} = 2\Ds/v,
\label{deBM}
\eeq

for a linear motion of mean velocity $v$, or by the thermal de Broglie length 

\beq
\lambda_{th}=2\Ds/ \langle v^2 \rangle^{1/2},
\label{deBT}
\eeq

for a medium or an ensemble of particles.

Combining these ideas with the process of decoherence (Eq.\ref{decoherence}), we conclude that diffusion processes play two key roles in macroscopic quantum processes.  In the first case, $D$ and $\rho$ collectively drive fractal structures in a scale dependent process, with molecular, $nm$ and cell scale fractal structures determining the value of $\widehat V$ and the emergent MQP

\beq
Q  = -2{\Ds}^2 \, \frac{\Delta \sqrt{\rho}}{\sqrt{\rho}}.
\eeq

The subsequent emergence of ordered structures are in turn influenced by the external diffusive force, which competes with internal macroscopic quantum forces in the decoherence process.  

When considering quantized morphologies, if internal forces (Eq.\ref{quantpot})) are insufficient to maintain a coherent structure against environmental perturbation, we see decoherence associated with collapse of the complex velocity field $\widehat V$ to its pointer states.  This is reflected at macroscopic scales in the emergence of fractal structures, e.g., a fern (a fractal leaf), which still exhibits long range order, its $D_F$ being determined by the relative strength of the residual field, a concept not incompatible with Pietak \cite{Pietak2011}, who suggested that electromagnetic fields govern leaf structure. 

Beyond a critical point in the decline of charge density, charge will be insufficient to support the emergence of a long range fractal architecture, i.e., long range fluctuations $\Ds$ disappear, along with the complex velocity field $\widehat V$.  This step equates to full macroscopic quantum decoherence of the field. However, in most instances, charge is sufficient to disrupt the formation of a crystal lattice, leading to the emergence of disordered tumour-like structures, where external diffusive forces (Eq.\ref{diffpot}) dictate morphology.

The dual role associated with diffusive type processes offers a new fundamental insight into the interplay between thermodynamics and macroscopic quantum processes and their role in defining ordered structure at a range of different scales.  

\begin{figure}[!ht]
\begin{center}
\includegraphics[width=4cm]{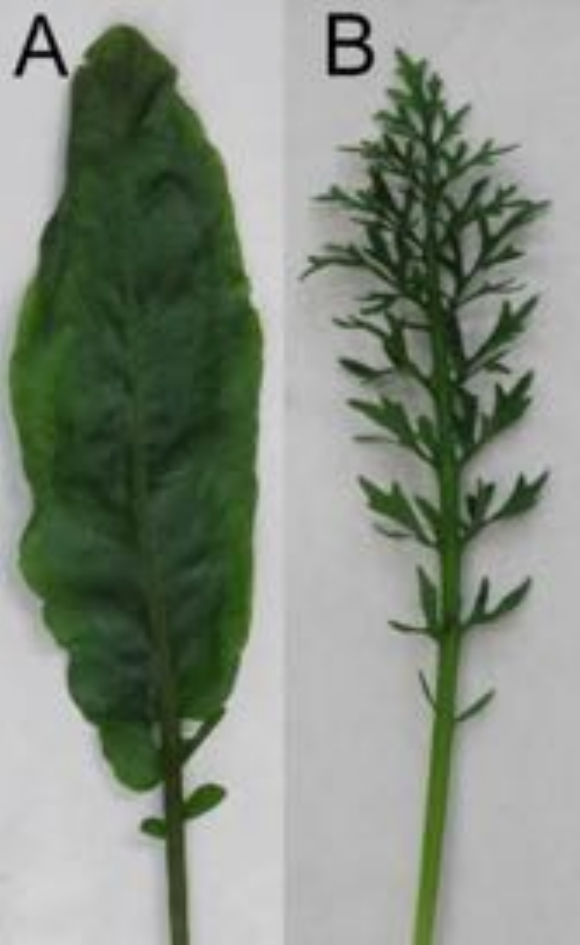} %
\caption{\small{Morphology of {\it{Rorippa aquatica}} leaves at $30^oC$ (A) and at $20^oC$ (B),}} reproduced from Nakamasu {\it{et al}} \cite{Nakamasu2014}. 
\label{Leaves}
\end{center}
\end{figure}

Based on the experimental work we have presented, it seems clear that changes in charge density and temperature, alongside other environmental conditions, have a critical role to play in the emergence of plant structure.  As suggested for mesoscale assemblies of cellulose nano-fibres in the cell wall (Section 4.2.4.1), within an evolutionary context, evidence suggests that genes encode mechanisms to control and maintain levels of charge density through transcription of the genetic code into proteins and protein complexes interacting with external sources of charge ($CO_2$ based generic protonation, phosphorylation etc).  The result is a range of macroscopic quantum processes leading to the emergence of the wide range of structures (e.g., cell, stems, branches, leaves, buds, flowers, seeds, pollen, fruits, pods) that we find in plants.  

Genetic control means that many of these processes are to an extent independent of environment.  This is supported by numerous examples of plants successfully introduced to climates different from those in which they evolved. Under most circumstances we do not see a radical change in plant morphology within a new environment.  However,  a number of plants are able to adapt their morphology in a significant way, in response to a changing environment  (phenotype plasticity), suggesting that the genome to metabalome translation process is more open to environmental influences.   In a recent study by Nakamasu {\it{et al}} \cite{Nakamasu2014}, changes in the morphology of leaves of North American Lake cress ({\it{Rorippa aquatica}}) with temperature (heterophylly) were reported.  Fig~\ref{Leaves}A illustrates a coherent (simple) leaf morphology in a plant grown at $30^oC$.  This contrasts with Fig~\ref{Leaves}B illustrating a branched (fractal) leaf grown at  $20^oC$.  These structures have their equivalence in Figure~\ref{Leaf} and the fractal structure observed in Figure~\ref{Plant structures}f. Further work on this species \cite{Nakayama2014} shows an even wider variation in fractal dimension over the temperature range from $15^oC$ to $30^oC$, which is expressed as a progressive morphogenic gradient, which we speculate may in part (directly or indirectly) be related to reduced levels of charge density as $CO_2$ concentration declines with temperature.  This temperature/morphology relationship was observed in both terrestrial and submerged plants.  However, for the same temperature, submerged plants expressed a far higher level of branching and fractal dimension.  A phenomenon which we suggest could also be attributable to lower levels of $CO_2$ availability in the aquatic environment.

Summarizing key findings from this second study \cite{Nakayama2014}, it was found that the concentration of endogenous gibberellin (GA) significantly increased as temperature increased from $20^oC$ to $25^oC$. The subsequent application of GA and uniconazole (a GA biosynthesis inhibitor) confirmed that GA leads to simplified leaf forms at $20^oC$ and $25^oC$ compared with a control, whilst uniconazole treatment increased leaf complexity.  The fact that higher concentrations of a charged molecule (GA) are produced at higher temperatures, sits well with our theory that as charge density increases, we expect a transition from structures with high fractal dimension to increased levels of spatial coherence.  

Since GA biosynthesis is regulated by KNOX1 genes \cite{Nakayama2014}, it was concluded that the KNOX-GA module is a key factor in the development of leaf morphology in {\it{R. aquatica}}. However, the overall picture is more complex.  A global analysis of the transcriptome linked the expression of 630 genes with dissected (fractal) leaves and 471 genes with simple leaves.  

The top 200 genes upregulated in the dissected leaf condition largely overlapped with genes upregulated in response to cold treatment, whilst those upregulated in the simple leaf condition overlapped with genes that are downregulated in response to cold treatment. To this complexity we can add the possible influence of $CO_2$ availability.  As suggested with respect to the work reported by Derbyshire {\it{et al}} \cite{Derbyshire2015}, there is a strong possibility that protons released through $CO_2$ dissolution can affect overall charge density during gene transcription and/or post-translational modification of proteins to influence molecular assembly.  Such an outcome would correlate with the effects of GA concentration, expressed through the KNOX-GA module and the hundreds of additional genes playing a role in the final expression of leaf morphology.

On a different level, the genes upregulated in the dissected leaf condition also overlapped with those that respond to changes in light intensity suggesting a further potential influence on leaf morphology. This was confirmed in growth experiments at various light intensities at constant temperature ($20^oC$).  There was a striking increase in leaf fractal dimension with increasing light intensity.  This fits well with our hypothesis that we are dealing with a macroscopic quantum system, with increasing fractal dimension reflecting photon induced decoherence as light intensity increases. 

The work by Nakayama {\it{et al}} \cite{Nakayama2014} represents new insights into the genetic control of leaf morphology in {\it{R. aquatica}}.  However, it leaves open the fundamental mechanisms at work in defining structure.  The concept of macroscopic quantum processes dictated by the genome, through translation into the proteome and its interaction with the environment, offers new insight into a global mechanism by which the genetic code is finally expressed in the form of plant structure.

\subsection*{4.3. Summing up}
Taking into account the effects of scale described in the theory of scale relativity, this paper addresses a key question related to the debate around irreversible laws linked to determinism and probablistic descriptions in physics.  Within this new theoretical framework, the laws of physics can jump from reversible to irreversible behavior and reciprocally, dependent on the scale.  For example, at the very smallest scales in nature we see a fundamental irreversibility under the reflection ${|dt|}\to-{|dt|}$, which is at the origin of the complex number representation of quantum mechanics (see Section 2.2). When these two real irreversible processes (which can be described by real path integrals) are combined into a complex one (a complex path integral), the new complex description (leading to the wave function) becomes reversible.  Decoherence of the wave packet introduces a new irreversibility in physics. Complete decoherence leads to classical physics, which is at first reversible. However, these classical laws, when considered on long time scales (larger than the Lyapunov timescale), dictated by the second law of thermodynamics, become chaotic, i.e. unpredictable and irreversible again. One of the main propositions of scale relativity \cite{Nottale1993,Nottale2011}, tested in the present paper, has been that on even longer time scales (initially predicted at >10 to 20 Lyapunov timescales), this basic irreversibility found at the scale of development of chaos becomes the seed for a new macroscopic quantum theory\footnote{which we have shown shares many features with standard QM}, whose equations are again reversible in terms of complex wave functions.  

As outlined in detail in the paper, the conditions required for the emergence of such a macroscopic quantum system are very specific.   Key amongst these is that charge density is sufficient to generate a charge induced interconnecting geometric network (along the lines illustrated in Fig~\ref{charges}), which at a critical percolation threshold leads to the emergence of a macroscopic quantum potential.  In theory this process can repeat itself over a very broad range of different scales. This principle is exemplified in the emergence of the cell, which as the `quanta of life' then repeats the process of reversibility/irreversibility, leading to the emergence of ordered multicellular structures.  However, the principle is also applicable at smaller scales.  Depending on specific conditions, a very diverse range of structures at different scales can emerge within a single cell.  Examples range from protein complexes \cite{Vattay2015}, nano-structures ($\approx$ 5-1000 $nm$  in Fig's \ref{spheres}B, \ref{Nanocrystals}, and \ref{Arabidopsis}B), to a multitude of cell organelles.  

In a generic process analogous to that suggested by Vattay {\it{et al}} \cite{Vattay2015} for proteins, we propose that the emergence of each of the many different types of potential structures is the result of a unique set of physical conditions dictated by both external and local conditions (defined by the cells genetic code), which have been selected for and refined during the evolutionary process.  

Following these principles, it should be possible to model each of these different structures with a macroscopic Schr\"odinger equation.  The challenge in achieving this is to correctly identify the detailed set of physical conditions leading to the emergence of each structure.  However, such an approach may prove extremely difficult in anything but the simplest biological processes.  An alternative starting point is to vary the external field/limiting conditions and or symmetries (the three elements which determine the solutions of the Schrodinger equation) and then see the morphology changing accordingly\footnote{A flower-like structure such as Fig~\ref{Morphogenesis} represents one of the simplest solutions (free case with no external field)}.  This approach has the potential to offer completely new coarse-grained, testable insights into the conditions that lead to the structures that emerge, generating signposts to support the process of decoding the complex role of the genome, transcriptome, proteome and metabalome in establishing these initial conditions.

We stress the point that above or below a critical level of charge density, the emergence of a macroscopic quantum system is not a foregone conclusion.  As  shown in Fig~\ref{cells}a, very low levels of charge density lead to crystalline structures.  As levels of charge increase beyond this point we see the emergence of a range of charge induced fractal networks (including tumors) which remain disordered and below the threshold for quantum criticality.  By contrast, at very high levels of charge density,  molecular repulsion leads to breakdown of structures and molecular disorder.  Examples to illustrate this process include dissolution (e.g. acid hydrolysis) or vaporization (e.g. plasma ashing\footnote{we note however that ionized gases can at an appropriate scale and charge density lead to a larger scale macroscopic quantum state and the emergence of long range order.}). 

As stated at the outset of the paper (Section 1), whilst the work presented here has focused specifically on structure in plants as a means to test and validate the basic principles we have described, it is clear that they are are not exclusive to plants, but generally applicable to all living and non living systems.  This is exemplified by the reference to work on planetary nebulae \cite{da Rocha2003a,da Rocha2003b} in Section 4.2.3 as well as many other astrophysical phenomena \cite{Nottale2011} where the transition from irreversibility (chaos) to reversibility (order) repeats itself.

This point highlights a far reaching implication of the theory of scale relativity.  There is a tendency to consider the cell as a relatively small object.  However, as noted in earlier work \cite{Nottale2011}, when taking the log of all scales in nature (the Planck scale $10^{-33}cm$ to the Cosmological scale $10^{28}cm$) then the cell, at $3$x$10^{-3}cm$ actually sits in the middle of this range of scales.  Within this context, the cell can be regarded as a (relatively) vast object in space, putting the analogies with astrophysical phenomena into a new perspective.

Within this much broader context, we emphasize the fact that, although the new macroscopic quantum theory shares many structures and methods with standard QM, it is NOT "Quantum Mechanics", since it is not based on the universal Planck constant $\hbar$, but on a parameter that may vary from one system to the other.  For example, in the application to astrophysics, this parameter comes under the equivalence principle (implying that the inertial mass disappears from the equations).  This endows astrophysical macroscopic quantum theory with specific characters different from standard QM (e.g., it is not energy $E$ or momentum $p$ which are quantized, but $E/m=v^2$ and $p/m=v$, which involves a quantization of velocity itself). In a similar way, we may expect that the application of this macroscopic quantum theory to living systems will endow it with specific biological characters, to be studied in more detail in future works.

On a final point, the findings outlined in this work have important implications, not just for our understanding of the emergence of structure, but also examples of collective behaviour in condensed matter such as coherent random lasing, high temperature superconductivity and photosynthesis, whose detailed mechanisms have to date, proved challenging to elucidate.  Within this context, we suggest that the work presented here offers a number of signposts and tools to assist future workers in deconstructing and understanding some of the wide range of phenomena (structures and functions) we have identified in the paper.  We also suggest that the new insights we have outlined have the potential to play an important role in supporting the development of new materials and  harness the potential applicability of macroscopic quantum phenomena in a range of different technologies.

\subsection*{\centering{5. Conclusions}}

We have developed an evidence based theory for the emergence of a range of structures associated with plants (e.g. crystals, tumours, ferns, fungi, stems, seeds, flowers pods, fruits, cells) through growth processes dictated by the presence of a fractal network of charges. This leads to a charge induced geometric landscape, which dictates macroscopic fluctuations ($\Ds$) and the emergence of macroscopic quantum potentials ($Q_N$ and $Q_{POL}$), which in turn drive macroscopic quantum processes and the emergence of different types of structure.  

At a range of plant scales, the strength of the dominant MQP (Eq \ref{eq.QMG}) and associated forces are a function of charge density $\rho$ and the velocity field $\widehat V$ of a fluid in potential motion, described in terms of a macroscopic complex wave function $\psi=\sqrt{\rho} {\times e^{iA/\hbar}}$.  The strength of the MQP determines the type of structure, e.g. dendritic (fern-like) structures or spherical (cell-like) structures.  Whilst spherical structures emerge at the highest levels of charge, dendritic structures signify the start of a macroscopic decoherence process to the roots (pointer states) of a charge induced fractal network `the field', which as we shall show in more detail future work, has its geometric equivalence in standard QM.

The macroscopic quantum processes we observe in plants take the form of a two component [coherent (boson)-classical (fermion)] system, which has its equivalence in CRL and HTSC \cite{Turner2015}.  Only bosons remain coherent at macroscopic scales.  The assembly of molecules (fermions) into plants operates within the framework of macroscopic fluctuations within a coherent bosonic field $\widehat V$, acting as a structuring force in competition with exterior potentials $\phi$ ($H_E$).  Whilst the system is only partially coherent, many of the phenomena associated with standard quantum theory are recovered, including quantization, non-dissipation, self-organization, confinement, structuration conditioned by the environment, environmental fluctuations leading to macroscopic quantum decoherence and evolutionary time described by the time dependent Schr\"odinger equation, which describes models of bifurcation and duplication.  This work provides a strong case for the existence of a geometrically derived quintessence-like behaviour \cite{Witten2002}, with macroscopic quantum potentials and associated forces having their equivalence in standard QM and gravitational forces in general relativity.  

As levels of charge decrease to a point where they are insufficient to form a MQP, we see the disappearance of macroscopic quantum behaviour.  This is reflected in the first instance in structures driven by pure dissipative processes, resulting in tumour-like structures.  At yet lower levels of charge, we see the elimination of charge-induced disruption during molecular assembly, leading to the emergence of nematic order and crystalline structures.

In conclusion, the presence or absence of macroscopic quantum forces is dictated by charge and a combination of dissipative and quantum systems.  At one level, dissipative systems (environmental fluctuations) are critical to molecular assembly and the emergence of a fractal network of charges, which drive the emergence of a MQP and ordered structures.  At the same time, the emergent structure is dependent on the relative contributions of charge density and competing environmental fluctuations.  These two processes are therefore inextricably linked through both synergistic and competing relationships, dictating molecular assembly into different structures.

When considering assembly mechanisms within real plants, a brief case study \cite{Derbyshire2015} offers important insights, indicating that structure is expressed at a wide range of points and scales within the plant through transcription of genetic code into charged macromolecules (proteins and protein complexes).  A second set of case studies \cite{Nakamasu2014,Nakayama2014} suggests that external sources of charge such as atmospheric $CO_2$ may also interact with the genetically coded array of proteins to impact on final structure.  

As mentioned in Section 4.2.2, evidence for macroscopic quantum coherence in plants has been reported in photosynthetic systems \cite{Engel2007,Collini2010}.  If the theory presented in this paper is correct, then these types of observations represent the tip of the iceberg.   A key challenge for future work lies in understanding in more detail how genomic, proteomic, transcriptomic, and metabolomic information translates into charge density distribution and the mechanisms by which the genes finally express themselves in plant structure at different scales within the organism.

As stated at the outset, the aim of this work was to test and validate some of the basic physical principles and theories proposed in earlier work \cite{Auffray2008,Nottale2008}, using a basic set of case studies focused on plants.  We suggest that the new experimental evidence and theory proposed in the present paper contributes an important first step in meeting this objective, offering a more detailed insight into the mechanisms at work and establishing some basic principles that are more generally applicable in all living systems.  

At a broader level, this work offers new insight into evolutionary processes in structural biology, with selection at any point in time, being made from a wide range of spontaneously emerging potential structures (dependent on conditions), which offer advantage for a specific organism. This is valid for both the emergence of structures from a prebiotic medium and the wide range of different plant structures we see today.  

As a final point, it should be clear from this work that biological processes, structures and systems are not in any way priviliged.  They can rather be viewed as a extremely large set of very complex, interacting systems, which explains the slow progress in deconstructing these processes to their individual component mechanisms.

\subsection*{\centering{6. Future work}}

In the present work we have given just a few examples of possible emergent structures using $CO_2$ as the source of charge density.  It represents only a first step in developing an understanding the vast and diverse range of processes and mechanisms at work in living systems.  Future work will focus on more detailed theory, modelling and controlled experimental studies to determine the impact of atomic/molecular structure (including biopolymers), levels of charge, pH and temperature on emerging structures.  As part of this process we will consider the possibility to mimic biological systems through the use of protein complexes to more precisely control the emergence of specific structures.

The objective - to develop an improved understanding of how to manipulate molecular and $nm$-scale particles into different structures and the development of new materials from first principles.  Examples include:

\begin{itemize}
\item the development of tuneable fractal systems for a range of materials in applications, which require macroscopic quantum coherence, including HTSC, CRL and quantum computing systems.  For more detailed proposals on HTSC see \cite{Turner2016}.

\item the development of cell duplication processes leading to a new multi-scale `cellular' approach to the development of materials with different structures, which mimic biological systems.

\end{itemize}

Work should also include studies on emergence of living structures from prebiotic media. Success in this area may improve our understanding of processes involved in the origins of life.  

At a different level, multidisciplinary work to understand evolutionary processes in established plants could also prove beneficial.  Working within the limits of biological systems, future work should consider environmental impacts on structure through more extensive trials in heterophylic plants, including studies to independently vary temperature and levels of $CO_2$ concentration under controlled conditions.  

A multidisciplinary approach is also required to target a more detailed and fundamental understanding of the role of the genome in setting the internal conditions within plants (in the first instance) that control structure at different scales.  This could have wide ranging applications in evolutionary biology and plant breeding, including an improved understanding of both past and future adaptive responses.  As an example, success in this area could contribute to our understanding of the impact of past and future climate change (temperature and $CO_2$ concentration) on different species and where appropriate (e.g. key agricultural species), support the identification of genetic variants most adaptable to different environments including future climate change.

\subsection*{Acknowledgements}

The authors would like to thank John Zhao for his assistance in carrying out the experimental work described in section 4.  We also wish to thank Akiko Nakamasu {\it{et al}}. \cite{Nakamasu2014} for granting permission to use Fig~\ref{Leaves} extracted from their paper.  Thanks also go to Edouard Pesquet for his very useful discussions and insights into the genetic control of structure in plants.  Finally we would like to acknowledge the role of the COST office, and in particular COST Action FP1105 for facilitating the collaboration that led to this work.

\subsection*{\centering{7. Supplementary material}}

The supplementary material contained in this section is provided as a quick reference for key elements of the theory.  For more detailed information the reader is referred to references contained in the text.  For the most up to date and in depth analysis we refer the reader to reference \cite{Nottale2011}.

\subsection*{7.1. From Newton to the Schr\"odinger equation}
\label{7.1}

In this section we outline the derivation of the Schr\"odinger equation through a process analogous with general relativity where after expansion of the covariant derivative, the free-form motion equations can be transformed into a Newtonian equation in which a generalized force emerges, of which the Newton gravitational force is an approximation.  Following the same principle, the covariance induced by scale effects leads to a transformation of the equation of motion, which, as we demonstrate through a number of steps, becomes after integration, the Schr\"odinger equation.  In the construction of this approach we note that whilst equations take a classical form, this form is applied to non differentiable geometry, so that the result is no longer classical.

\subsection*{7.1.1. Momentum}

Due to the complex nature of the velocity field $\widehat V$ (Eq.~\ref{compvelfield}), the classic equation $p=mv,$ can be generalized to its complex representation \cite{Nottale2011}.  
\beq
{\widehat P} = m {\widehat V},
\label{eq.15}
\eeq
so that the complex velocity field $\widehat V$ is potential (irrotational), given by the gradient of the complex action,
\beq
{\widehat V} = \frac{ \nabla {\widehat S}}{m}.
\label{eq.cvf}
\eeq
We now introduce a complex function $\psi$ identifiable with a wave function or state function, which is another expression for the complex action $\widehat S$,
\beq
\psi = e^{i{\widehat S}/S_0}.
\label{eq.psi}
\eeq
The factor $S_0$ has the dimension of an action, i.e. of an angular momentum with $S_0=\hbar$ in the case of standard QM, where $\hbar$ is a geometric property of the fractal space, defined through the fractal fluctuations as $\hbar=2m{\widetilde D}=m\langle{d\xi^2}\rangle/dt$.

The next step consists of making a change of variable in which one connects the complex velocity field Eq.~(\ref{eq.cvf}), to a wave function, $\psi$ where $\ln\psi$ plays the role of a velocity potential according to the relation
\beq
{\widehat V} = - i \, \frac{S_0}{m} \, \nabla (\ln \psi).
\label{eq.vfwf}
\eeq
The complex momentum Eq.~(\ref{eq.15}) may now be written under the form
\beq
{\widehat P}= - i \,S_0 \, \nabla (\ln \psi),
\label{eq.19}
\eeq
i.e. 
\beq
{\widehat P} \psi= - i \,S_0 \, \nabla \psi.
\label{eq.20}
\eeq
In the case of standard QM ($S_0=\hbar$), this relation reads ${\widehat P} \psi= - i \,\hbar \, \nabla \psi$, i.e. it is a derivation of the principle of correspondence for momentum, $p \to -i\, \hbar\, \nabla$, where the real part of the complex momentum $\widehat P$ is, in the classical limit, the classical momentum $p$.  The `correspondence' is therefore understood as between the real part of a complex quantity and an operator acting on the function $\psi$.  However, thanks to the introduction of the complex momentum of the geodesic fluid, it is no longer a mere correspondence, it has become a genuine equality.
The same follows for angular momentum $L=rp$, which can also be generalized to the complex representation 
\beq
{\widehat L} \psi =  - i \,S_0 \; r \times \nabla \psi,
\label{eq.21}
\eeq
so that we recover, in the standard quantum case $S_0=\hbar$, the correspondence principle for angular momentum, which again emerges as an equality.

\subsection*{7.1.2. Remarkable identity}

We now write the fundamental equation of dynamics Eq.~(\ref{eq.dyn}) in terms of the new quantity $\psi$. 
\beq
i S_0 \frac{\widehat{d}}{dt}(\nabla \ln \psi) = \nabla \phi.
\label{eq.22}
\eeq
We note that $\widehat{d}$ and $\nabla$ do not commute.  However, as we shall see in what follows ${\widehat{d}}(\nabla \ln \psi)/dt$, is a gradient in the general case.
  
Replacing $\widehat{d}/{dt}$  by its expression, given by Eq.~(\ref{comptimderiv}), yields
\beq
\nabla   \phi  =  i S_0 \left(\frac{\partial}{\partial t} + {\widehat V}. 
\nabla - i {\widetilde D} \Delta\right) (\nabla \ln \psi),
\label{eq.23}   
\eeq
and replacing once again ${\widehat V}$ by its expression in Eq.~(\ref{eq.vfwf}), we obtain 
\beq
\nabla   \phi  =   i S_0 \left\{ \frac{\partial }{\partial t} \nabla   
\ln\psi   - i \left[  \frac{S_0}{m} (\nabla   \ln\psi  . \nabla   )
(\nabla   \ln\psi ) + {\widetilde D} \Delta (\nabla   \ln\psi )\right]\right\} .
\label{eq.nablaphi}
\eeq
Consider now the identity \cite{Nottale1993}  
\beq
(\nabla \ln f)^{2} + \Delta \ln f =\frac{\Delta f}{f} \; ,
\label{eq.25}
\eeq
which proceeds from the following tensorial derivation
\begin{eqnarray}
\partial_{\mu} \partial^{\mu} \ln f +\partial_{\mu} \ln f \partial^{\mu} 
\ln f &=& \partial_{\mu} \frac{\partial^{\mu} f}{f}+\frac{\partial_{\mu} f}
{f}\frac{\partial^{\mu} f}{f} \nonumber \\
 &=& \frac{f \partial_{\mu} \partial^{\mu} f - 
\partial_{\mu} f \partial^{\mu} f}{f^{2}}+\frac{ \partial_{\mu} f 
\partial^{\mu} f}{f^{2}} \nonumber \\
&=& \frac{\partial_{\mu} \partial^{\mu} f}{f} \; . 
\label{eq.26}
\end{eqnarray}
When we apply this identity to $\psi$ and take its gradient, we obtain
\beq
\nabla\left(\frac{\Delta \psi}{\psi}\right)=\nabla[(\nabla \ln \psi)^{2} + 
\Delta \ln \psi] .
\label{eq.27}
\eeq
The second term on the right-hand side of this expression can be transformed, 
using the fact that $\nabla$ and $\Delta$ commute, i.e.,
\beq
\nabla \Delta =\Delta \nabla . 
\label{eq.28}
\eeq
The first term can also be transformed thanks to another identity,
\beq
\nabla (\nabla f)^{2}=2 (\nabla f . \nabla) (\nabla f) ,
\label{eq.29}
\end{equation}
that we apply to $f=\ln \psi$. We finally obtain \cite{Nottale1993}
\beq
\nabla\left(\frac{\Delta \psi}{\psi}\right)= 2 (\nabla \ln\psi . \nabla )
(\nabla \ln \psi )  + \Delta (\nabla \ln \psi).
\label{eq.30}
\eeq
This identity can be still generalized thanks to the fact that $\psi$ appears only through its logarithm in the right-hand side of the above equation. By replacing  $\psi$ with $\psi^{\alpha}$, we obtain the general remarkable identity \cite{Nottale2008b}
\beq
\frac{1}{\alpha} \; \nabla\left(\frac{\Delta  \psi^{\alpha}}{\psi ^{\alpha}}\right)= 2\alpha \, (\nabla \ln \psi . \nabla )(\nabla \ln \psi) + \Delta (\nabla \ln \psi).
\label{eq.RI}
\eeq

\subsection*{7.1.3. The Schr\"odinger equation}
We recognize in the right-hand side of Eq.~(\ref{eq.RI}) the two terms of Eq.~(\ref{eq.nablaphi}), which were respectively in factor of $S_0$ and $\widetilde D$.  Therefore, by writing the above remarkable identity in the case
\beq
\alpha= \frac{S_0}{2 m {\widetilde D}},
\label{eq.alpha}
\eeq
the whole motion equation becomes a gradient,
\beq
\nabla   \phi  =  2m  {\widetilde D}\left[ i\frac{\partial }{\partial t} \nabla   
\ln\psi^\alpha  +  {\widetilde D}   \nabla\left(\frac{\Delta  \psi^{\alpha}}{\psi ^{\alpha}} \right)\right],
\label{eq.33}
\eeq
and it can therefore be generally integrated, in terms of the new function
\beq
\psi^{\alpha}=  (e^{i{\widehat S}/S_0})^\alpha=e^{i{\widehat S}/2 m {\widetilde D}}.
\label{eq.34}
\eeq
which is more general than in standard QM, for which $S_0=\hbar=2m\widetilde D$.  Eq.~(\ref{eq.alpha}) is actually a generalization of the Compton relation.  This means that the function $\psi$ becomes a wave function only provided it comes with a Compton-de Broglie relation, a result which is  naturally achieved here.  Without this relation, the equation of motion would remain of third order, with no general prime integral.

The simplification brought by this relation means that several complicated terms are compacted into a simple one and that the final remaining term is a gradient, which means that the fundamental equation of dynamics can now be integrated in a universal way.  The function $\psi$ in Eq.~(\ref{eq.psi}) is therefore finally defined as
\beq
\psi = e^{i{\widehat S}/2m{\widetilde D}},
\label{eq.35}
\eeq
which is a solution of the fundamental equation of dynamics, Eq.~(\ref{eq.dyn}), 
which now takes the form
\beq
\frac{\widehat{d}}{dt} {\widehat V} = -2 {\widetilde D} \nabla \left(i \frac{\partial}
{\partial t} \ln \psi + {\widetilde D} \frac{\Delta \psi}{\psi}\right) = 
-\frac{\nabla \phi}{m}.
\label{eq.36}
\eeq	
Using the fact that $d ln\psi=d\psi/\psi$, the full equation becomes a gradient,
\beq
\nabla\left[\frac{\phi}{m}-2 {\widetilde D} \nabla \left(\frac{i \partial\psi/\partial t+\widetilde D\Delta\psi}{\psi}\right)\right] = 0.
\label{eq.37}
\eeq	
Integrating this equation finally yields a generalized Schr\"odinger equation, which is Eq.~(\ref{eq.Schrodinger-diff}) in the main text.
\beq
{\widetilde D}^2 \Delta \psi + i {\widetilde D} \frac{\partial}{\partial t} \psi - 
\frac{\phi}{2m}\psi = 0,
\label{eq.Schrodinger}
\eeq

\subsection*{7.2. Fluid representation with a macroscopic quantum potential}

In this section we demonstrate the fundamental meaning of the wave function as a wave of probability, and that the geodesic equation can take not only a Schr\"odinger form, but also a fluid dynamics form with an added quantum potential.  We begin by writing the wave function under the form $\psi=\sqrt{P}\times{e}^{iA/\hbar}$, decomposing it in terms of a phase, defined as a dimensioned action $A$ and a modulus $P= |\psi|^2$, which gives the number density of virtual geodesics \cite{Nottale2011,Nottale2007}.  This function becomes naturally a density of probability.  The function $\psi$, being a solution of the Schr\"odinger equation and subjected to the Born postulate and to the Compton relation, therefore owns most of the properties of a wave function.

The complex velocity field $\widehat V$ Eq.~(\ref{compvelfield}) can be expressed in terms of the classical (real) part of the velocity field $V$ and of the number density of geodesics $P_N$, which as we have shown is equivalent to a probability density $P$ where
\beq
\Vsc= V- i \Ds \nabla \ln P.
\label{eq.42}
\eeq
The quantum covariant derivative operator thus reads
\beq
\frac{\dfr}{\d t}=\frac{\d}{\d t} + V. \nabla -i \Ds \: ( \nabla \ln P.\nabla+\Delta).
\label{eq.43}
\eeq
When we introduce an exterior scalar potential $\phi$, the fundamental equation of dynamics becomes
\beq
\l( \frac{\d}{\d t} + V. \nabla -i \Ds \: ( \nabla \ln P.\nabla+\Delta) \r) ( V- i \Ds \nabla \ln P)=-\frac{ \nabla \phi}{m}.
\label{eq.44}
\eeq
The imaginary part of this equation,
\beq
\Ds \: \l[(\nabla \ln P.\nabla+\Delta)V + \l( \frac{\d}{\d t} + V. \nabla\r)\nabla \ln P \r]=0,
\label{eq.45}
\eeq
takes, after some calculations, the following form
\beq
\nabla\l[\frac{1}{P}  \l( \frac{\d P}{\d t} + \text{div}(P V)\r)\r]=0,
\label{eq.46}
\eeq
which can finally be integrated in terms of a continuity equation:
\beq
\frac{\d P}{\d t} + \text{div} ( P V)=0.
\label{continuity}
\eeq
The real part,
\beq
\l( \frac{\d}{\d t} + V. \nabla\r) V=-\frac{ \nabla \phi}{m} +\Ds^2 \: ( \nabla \ln P.\nabla+\Delta) \nabla \ln P,
\label{eq.48}
\eeq
takes the form of a Euler equation, 
\beq
m \l( \frac{\d}{\d t} + V. \nabla\r) V=- \nabla \phi +2 m \Ds^2\nabla \l( \frac{\Delta\sqrt{P}}{\sqrt{P}}\r),
\label{Euler}
\eeq  
which describes a fluid subjected to an additional quantum potential $Q$ that depends on the probability density $P$
\beq
Q=-2 m \Ds^2 \: \frac{ \Delta \sqrt{P}}{\sqrt{P}}.
\label{Eulerquantpot}
\eeq

This approach is similar to the Madelung transformation \cite{Madelung1927}, but in a way that all its various elements make sense from first principles, instead of being postulated.  Since the Schr\"odinger equation is obtained as a reformulation of the geodesic equation, it is possible to go directly from the covariant equation of dynamics Eq.~(\ref{eq.dyn}) to the fluid mechanics equations without defining the wave function or passing through the Schr\"odinger equation.

\subsection*{{7.3. A fluid representation of a diffusive system. }}

Consider a classical diffusion process described by a Fokker-Planck equation,
\beq
\frac{\d P}{\d t}+ {\rm div}(P v)=D \Delta P,
\label{FP}
\eeq
where $D$ is the diffusion coefficient, $P$ the probability density distribution of the particles and $v(x,t)$ is their mean velocity.

When there is no global motion of the diffusing fluid or particles ($v=0$), the Fokker-Planck equation is reduced to the usual diffusion equation for the probability $P$,
\beq
\frac{\d P}{\d t}=D \Delta P.
\label{eq.52}
\eeq
Conversely, when the diffusion coefficient vanishes, the Fokker-Planck equation is reduced to the continuity equation,
\beq
\frac{\d P}{\d t}+ {\rm div}(P v)=0.
\label{eq.53}
\eeq
We now make a change of variable, where in the general case, $v$ and $D$ are a priori non vanishing, 
\beq
V=v-D \nabla \ln P.
\label{eq.54}
\eeq
We first prove that the new velocity field $V (x, y, z, t)$ is a solution of the standard continuity equation.  Taking the Fokker-Planck equation and replacing $V$ by its above expression, we find
\beq
\frac{\d P}{\d t} + {\rm div} (P V)=\{ D \Delta P-{\rm div}(P v)  \} +  {\rm div} (Pv)-D \,  {\rm div} (P\nabla \ln P).
\label{eq.55}
\eeq
Finally the various terms cancel each other and we obtain the continuity equation for the velocity field $V$,
\beq
\frac{\d P}{\d t} + {\rm div} (P V)=0.
\label{eq.56}
\eeq
Therefore the diffusion term has been absorbed in the re-definition of the velocity field.

We now consider a fluid-like description of the diffusing motion and determine the form of the Euler equation for the velocity field $V$.
As a first step we consider the case of vanishing mean velocity.

Let us calculate the total time derivative of the velocity field $V$, first in the simplified case $v = 0$
\beq 
\frac{d}{dt}V= \l( \frac{\d}{\d t} + V. \nabla\r)V=- D\,\frac{\d}{\d t} \nabla \ln P + D^2 (\nabla \ln P. \nabla) \nabla \ln P.
\label{57}
\eeq
Since ${\d}\nabla \ln P /{\d t}=\nabla {\d} \ln P /{\d t}=\nabla (P^{-1}{\d} P /{\d t})$, we can make use of the diffusion equation so that we obtain
\beq
\l( \frac{\d}{\d t} + V. \nabla\r)V=- D^2 \l[   \nabla \l( \frac{\Delta P}{P}\r)-(\nabla \ln P. \nabla) \nabla \ln P  \r].
\label{eq.58}
\eeq
In order to write this expression in a more compact form, we use the fundamental remarkable identity Eq.~(\ref{eq.34}), where $\psi=R$
\beq
\frac{1}{\alpha} \; \nabla\left(\frac{\Delta  R^{\alpha}}{R ^{\alpha}}\right)=   \Delta (\nabla \ln R)+2\alpha (\nabla \ln R . \nabla )
(\nabla \ln R ).
\label{59}
\eeq
By writing this remarkable identity for $R=P$ and $\alpha=1$, we can replace $  \nabla ({\Delta P}/{P})$ by $ \Delta (\nabla \ln P)+2 (\nabla \ln P . \nabla )\nabla \ln P$, so that Eq.~(\ref{eq.58}) becomes
\beq
\l( \frac{\d}{\d t} + V. \nabla\r)V=- D^2 \l\{   \Delta (\nabla \ln P)+(\nabla \ln P. \nabla) \nabla \ln P  \r\}.
\label{eq.60}
\eeq
The right-hand side of this equation comes again under the identity Eq.~(\ref{eq.58}), but now for $\alpha=1/2$. Therefore we finally obtain the following form for the Euler equation of the velocity field $V$, which describes a diffusive system:
\beq
\l( \frac{\d}{\d t} + V. \nabla\r)V=- 2 D^2 \, \nabla \l( \frac{\Delta \sqrt{P}}{\sqrt{P}}\r).
\label{DP}
\eeq

%%%%%%%%%%%%

 %%%%%%%%%%%

%%%%%%%%%%
\end{document}